\documentclass[a4paper,11pt]{article}
\pdfoutput=1 
\usepackage{jcappub} 

\usepackage[T1]{fontenc} 
\usepackage{physics}
\usepackage{multirow}
\usepackage[table,xcdraw]{xcolor}
\usepackage[skip=0.5\baselineskip]{caption}
\definecolor{mygray}{gray}{0.9}
\usepackage{aas_macros}
\usepackage{orcidlink}

\usepackage[nameinlink,noabbrev]{cleveref}
\crefname{equation}{Eq.}{Eqs.}
\crefname{section}{Section}{Sections}
\crefname{figure}{Fig.}{Figs.}
\crefname{table}{Table}{Tables}
\crefname{appendix}{Appendix}{Appendices}
\Crefname{figure}{Figure}{Figures}
\Crefname{equation}{Equation}{Equations}
\Crefname{section}{Section}{Sections}
\Crefname{table}{Table}{Tables}

\title{\boldmath Analytical and EZmock covariance validation for the DESI 2024 results}

\author[1,13]{{D.~Forero-Sánchez}\orcidlink{0000-0001-5957-332X},}
\author[2]{{M.~Rashkovetskyi}\orcidlink{0000-0001-7144-2349},}
\author[3]{{O.~Alves},}
\author[4]{{A.~de~Mattia}\orcidlink{0000-0003-0920-2947},}
\author[5]{{N.~Padmanabhan},}
\author[6]{{H.~Seo}\orcidlink{0000-0002-6588-3508},}
\author[7]{{S.~Nadathur}\orcidlink{0000-0001-9070-3102},}
\author[8,9,10]{{A.~J.~Ross}\orcidlink{0000-0002-7522-9083},}
\author[11,12,13]{{H.~Gil-Mar\'in}\orcidlink{0000-0003-0265-6217},}
\author[14]{{P.~Zarrouk}\orcidlink{0000-0002-7305-9578},}
\author[1]{{J.~Yu}\orcidlink{0009-0001-7217-8006},}
\author[15]{{Z.~Ding}\orcidlink{0000-0002-3369-3718},}
\author[16,3]{{U.~Andrade}\orcidlink{0000-0002-4118-8236},}
\author[5]{{X.~Chen}\orcidlink{0000-0003-3456-0957},}
\author[2,17]{{C.~Garcia-Quintero}\orcidlink{0000-0003-1481-4294},}
\author[18]{{J.~Mena-Fern\'andez}\orcidlink{0000-0001-9497-7266},}
\author[19]{{S.~Ahlen}\orcidlink{0000-0001-6098-7247},}
\author[20]{{D.~Bianchi}\orcidlink{0000-0001-9712-0006},}
\author[21]{{D.~Brooks},}
\author[4]{{E.~Burtin},}
\author[22]{{E.~Chaussidon}\orcidlink{0000-0001-8996-4874},}
\author[22]{{T.~Claybaugh},}
\author[23]{{S.~Cole}\orcidlink{0000-0002-5954-7903},}
\author[24]{{A.~de la Macorra}\orcidlink{0000-0002-1769-1640},}
\author[24]{{M.~Enriquez-Vargas},}
\author[12,7,25]{{E.~Gaztañaga},}
\author[26]{{G.~Gutierrez},}
\author[8,27,10]{{K.~Honscheid}\orcidlink{0000-0002-6550-2023},}
\author[28]{{C.~Howlett}\orcidlink{0000-0002-1081-9410},}
\author[22]{{T.~Kisner}\orcidlink{0000-0003-3510-7134},}
\author[22]{{M.~Landriau}\orcidlink{0000-0003-1838-8528},}
\author[14]{{L.~Le~Guillou}\orcidlink{0000-0001-7178-8868},}
\author[22]{{M.~E.~Levi}\orcidlink{0000-0003-1887-1018},}
\author[29,30]{{R.~Miquel},}
\author[31]{{J.~Moustakas}\orcidlink{0000-0002-2733-4559},}
\author[4,22]{{N.~Palanque-Delabrouille}\orcidlink{0000-0003-3188-784X},}
\author[32,33,34]{{W.~J.~Percival}\orcidlink{0000-0002-0644-5727},}
\author[35]{{I.~P\'erez-R\`afols}\orcidlink{0000-0001-6979-0125},}
\author[36]{{G.~Rossi},}
\author[37]{{E.~Sanchez}\orcidlink{0000-0002-9646-8198},}
\author[22]{{D.~Schlegel},}
\author[38,3]{{M.~Schubnell},}
\author[39]{{D.~Sprayberry},}
\author[3]{{G.~Tarl\'{e}}\orcidlink{0000-0003-1704-0781},}
\author[24]{{M.~Vargas-Maga\~na}\orcidlink{0000-0003-3841-1836},}
\author[39]{{B.~A.~Weaver},}
\author[40]{{H.~Zou}\orcidlink{0000-0002-6684-3997},}

\affiliation[1]{Institute of Physics, Laboratory of Astrophysics, \'{E}cole Polytechnique F\'{e}d\'{e}rale de Lausanne (EPFL), Observatoire de Sauverny, Chemin Pegasi 51, CH-1290 Versoix, Switzerland}
\affiliation[2]{Center for Astrophysics $|$ Harvard \& Smithsonian, 60 Garden Street, Cambridge, MA 02138, USA}
\affiliation[3]{University of Michigan, Ann Arbor, MI 48109, USA}
\affiliation[4]{IRFU, CEA, Universit\'{e} Paris-Saclay, F-91191 Gif-sur-Yvette, France}
\affiliation[5]{Physics Department, Yale University, P.O. Box 208120, New Haven, CT 06511, USA}
\affiliation[6]{Department of Physics \& Astronomy, Ohio University, Athens, OH 45701, USA}
\affiliation[7]{Institute of Cosmology and Gravitation, University of Portsmouth, Dennis Sciama Building, Portsmouth, PO1 3FX, UK}
\affiliation[8]{Center for Cosmology and AstroParticle Physics, The Ohio State University, 191 West Woodruff Avenue, Columbus, OH 43210, USA}
\affiliation[9]{Department of Astronomy, The Ohio State University, 4055 McPherson Laboratory, 140 W 18th Avenue, Columbus, OH 43210, USA}
\affiliation[10]{The Ohio State University, Columbus, 43210 OH, USA}
\affiliation[11]{Departament de F\'{\i}sica Qu\`{a}ntica i Astrof\'{\i}sica, Universitat de Barcelona, Mart\'{\i} i Franqu\`{e}s 1, E08028 Barcelona, Spain}
\affiliation[12]{Institut d'Estudis Espacials de Catalunya (IEEC), 08034 Barcelona, Spain}
\affiliation[13]{Institut de Ci\`encies del Cosmos (ICCUB), Universitat de Barcelona (UB), c. Mart\'i i Franqu\`es, 1, 08028 Barcelona, Spain.}
\affiliation[14]{Sorbonne Universit\'{e}, CNRS/IN2P3, Laboratoire de Physique Nucl\'{e}aire et de Hautes Energies (LPNHE), FR-75005 Paris, France}
\affiliation[15]{Department of Astronomy, School of Physics and Astronomy, Shanghai Jiao Tong University, Shanghai 200240, China}
\affiliation[16]{Leinweber Center for Theoretical Physics, University of Michigan, 450 Church Street, Ann Arbor, Michigan 48109-1040, USA}
\affiliation[17]{Department of Physics, The University of Texas at Dallas, Richardson, TX 75080, USA}
\affiliation[18]{Laboratoire de Physique Subatomique et de Cosmologie, 53 Avenue des Martyrs, 38000 Grenoble, France}
\affiliation[19]{Physics Dept., Boston University, 590 Commonwealth Avenue, Boston, MA 02215, USA}
\affiliation[20]{Dipartimento di Fisica ``Aldo Pontremoli'', Universit\`a degli Studi di Milano, Via Celoria 16, I-20133 Milano, Italy}
\affiliation[21]{Department of Physics \& Astronomy, University College London, Gower Street, London, WC1E 6BT, UK}
\affiliation[22]{Lawrence Berkeley National Laboratory, 1 Cyclotron Road, Berkeley, CA 94720, USA}
\affiliation[23]{Institute for Computational Cosmology, Department of Physics, Durham University, South Road, Durham DH1 3LE, UK}
\affiliation[24]{Instituto de F\'{\i}sica, Universidad Nacional Aut\'{o}noma de M\'{e}xico,  Cd. de M\'{e}xico  C.P. 04510,  M\'{e}xico}
\affiliation[25]{Institute of Space Sciences, ICE-CSIC, Campus UAB, Carrer de Can Magrans s/n, 08913 Bellaterra, Barcelona, Spain}
\affiliation[26]{Fermi National Accelerator Laboratory, PO Box 500, Batavia, IL 60510, USA}
\affiliation[27]{Department of Physics, The Ohio State University, 191 West Woodruff Avenue, Columbus, OH 43210, USA}
\affiliation[28]{School of Mathematics and Physics, University of Queensland, 4072, Australia}
\affiliation[29]{Instituci\'{o} Catalana de Recerca i Estudis Avan\c{c}ats, Passeig de Llu\'{\i}s Companys, 23, 08010 Barcelona, Spain}
\affiliation[30]{Institut de F\'{i}sica d’Altes Energies (IFAE), The Barcelona Institute of Science and Technology, Campus UAB, 08193 Bellaterra Barcelona, Spain}
\affiliation[31]{Department of Physics and Astronomy, Siena College, 515 Loudon Road, Loudonville, NY 12211, USA}
\affiliation[32]{Department of Physics and Astronomy, University of Waterloo, 200 University Ave W, Waterloo, ON N2L 3G1, Canada}
\affiliation[33]{Perimeter Institute for Theoretical Physics, 31 Caroline St. North, Waterloo, ON N2L 2Y5, Canada}
\affiliation[34]{Waterloo Centre for Astrophysics, University of Waterloo, 200 University Ave W, Waterloo, ON N2L 3G1, Canada}
\affiliation[35]{Departament de F\'isica, EEBE, Universitat Polit\`ecnica de Catalunya, c/Eduard Maristany 10, 08930 Barcelona, Spain}
\affiliation[36]{Department of Physics and Astronomy, Sejong University, Seoul, 143-747, Korea}
\affiliation[37]{CIEMAT, Avenida Complutense 40, E-28040 Madrid, Spain}
\affiliation[38]{Department of Physics, University of Michigan, Ann Arbor, MI 48109, USA}
\affiliation[39]{NSF NOIRLab, 950 N. Cherry Ave., Tucson, AZ 85719, USA}
\affiliation[40]{National Astronomical Observatories, Chinese Academy of Sciences, A20 Datun Rd., Chaoyang District, Beijing, 100012, P.R. China}

\emailAdd{daniel.forerosanchez@epfl.ch}

\newcommand{\Gpch}{\,h^{-1}\,{\rm Gpc}}
\newcommand{\Mpch}{\,h^{-1}\,{\rm Mpc}}
\newcommand{\hMpc}{\,h\,{\rm Mpc}^{-1}}
\newcommand{\rascalc}{{\sc RascalC}}
\newcommand{\thecov}{{\sc TheCov\,}}
\newcommand{\ezmock}{{\tt EZmocks\,}}

\abstract{The estimation of uncertainties in cosmological parameters is an important challenge in Large-Scale-Structure (LSS) analyses. For standard analyses such as Baryon Acoustic Oscillations (BAO) and Full-Shape two approaches are usually considered. First: analytical estimates of the covariance matrix use Gaussian approximations and (nonlinear) clustering measurements to estimate the matrix, which allows a relatively fast and computationally cheap way to generate matrices that adapt to an arbitrary clustering measurement. On the other hand, sample covariances are an empirical estimate of the matrix based on an ensemble of clustering measurements from fast and approximate simulations. While more computationally expensive due to the large amount of simulations and volume required, these allow us to take into account systematics that are impossible to model analytically. In this work we compare these two approaches in order to enable DESI’s key analyses. We find that the configuration space analytical estimate performs satisfactorily in BAO analyses and its flexibility in terms of input clustering makes it the fiducial choice for DESI's 2024 BAO analysis. On the contrary, the analytical computation of the covariance matrix in Fourier space does not reproduce the expected measurements in terms of Full-Shape analyses, which motivates the use of a corrected mock covariance for DESI's 2024 Full Shape analysis.}

\begin{document}
\maketitle
\flushbottom

\section{Introduction}
\label{sec:intro}
Over the last decades, observational cosmology has become a data-driven field of physics thanks to the huge leaps in our ability to observe the Universe through a multitude of observables. In this work, we are concerned with the observations of the distribution of galaxies in the Universe as observed by the Dark Energy Spectroscopic Instrument (DESI). DESI is expected to observe a field of 14~000 $\deg^2$, the largest so far, by measuring the precise position of 40 million objects from the near Universe at $z\sim 0.1$ until $z\sim 2$ \citep{Snowmass2013.Levi,DESI2016a.Science,DESI2016b.Instr,DESI2022.KP1.Instr,FocalPlane.Silber.2023,Corrector.Miller.2023,Spectro.Pipeline.Guy.2023,SurveyOps.Schlafly.2023,Levi2019, DESI2023a.KP1.SV,DESI2023b.KP1.EDR}, exceeding by far the observations of its predecessor the Sloan Digital Sky Survey (SDSS) \citep{Eisenstein2011, Dawson2013}. The main objective of these spectroscopic surveys is to observe the largest possible amount of matter tracer positions in the largest possible cosmological volume to compute precise clustering measurements where Baryon Acoustic Oscillations (BAO) are detected. The BAO analysis has been used for 20 years \cite{Cole05, Eisenstein05, Beutler11, Alam2021} to deduce cosmological information from galaxy clustering and is known to be a robust method to constrain the expansion history of the Universe. In addition, Full-Shape analyses allow us to constrain the growth of structure through cosmic history. 
DESI focuses on so-called “full-shape” analyses (e.g. \cite{Brieden21,DESI2024.V.KP5}) which extract not only the information on Redshift Space Distortions, but also extracts information on the shape of the clustering, therefore shown to be more informative over the standard RSD analysis used in eBOSS \citep{Maus23}.

In either case, observing the Universe at its largest scale presents an unavoidable challenge: we can only observe one universe. The absence of an ensemble of observable universes implies we cannot repeat our observations to estimate errors in our measurements and we are forced to rely on either simulated universes -- that we must make as realistic as possible -- or on analytical estimates of the covariances of our measurements. 

This paper aims to compare the analytical estimates of the covariance matrix against the ``true'' sample covariance matrix. We do this by generating analytical matrices based on the mock sample itself -- as opposed to generating them from the data --, thereby removing the assumption that the distribution of mocks is the same distribution underlying the data. 
In particular, we compare both techniques directly (i.e. the matrices themselves) and in terms of their impact on the BAO and the full shape fits of DESI. These studies allow us to detect any potential bias in the parameter or error estimates due to the covariance matrix choice. This work is part of the set of supporting papers around the DESI BAO \cite{DESI2024.III.KP4} and Full-Shape \cite{DESI2024.V.KP5} analyses using the DESI Data Release 1 (DR1) \cite{DESI2024.I.DR1}. Other relevant papers using DR1 include the clustering measurements from galaxies and quasars, \cite{DESI2024.II.KP3}, the Lyman-$\alpha$ forest results \cite{DESI2024.IV.KP6} as well as the cosmological results from BAO \cite{DESI2024.VI.KP7A} and full-shape \cite{DESI2024.VII.KP7B}.

This paper is organised as follows: in \cref{sec:methods} we briefly introduce the methods used in this comparison.
First regarding the fitting techniques and relevant cosmological parameters and second, the estimation of sample and analytical covariances in configuration and Fourier space. Then, \cref{sec:results} reports the results of our tests, with \cref{sec:conf-results} focusing on configuration-space-based measurements and BAO analyses whereas \cref{sec:fourier-results} focuses on Fourier-space measurements and Full-Shape analyses. Finally, we discuss and conclude in \cref{sec:discussion-conclusion}.

\section{Methods}
\label{sec:methods}
\subsection{Mock sample}
\label{sec:mocks}
This work relies on the existence of a large sample of mock Universes (or ``mocks'') that are realistic enough to reproduce the observed galaxy samples; which could then be assumed to have been drawn from the same underlying probability distribution as the mocks. However, generating a large sample of precise $N$-body mocks is a computational challenge, as DESI not only observed a much larger effective cosmological volume than its predecessors, but did so with significantly greater precision. To alleviate this, various approaches have emerged in which we trade precision in the gravitational evolution of the matter field and accuracy at small scales for computational efficiency. This trade-off enables the generation of a large number of mocks, which are required to build accurate mock-based covariance matrices. We refer the reader to refs.~\cite{Chuang2015, Lippich2019,Blot2019,Colavincenzo2019} for comprehensive studies of the various methods to generate these approximate mocks. 

Throughout this work, we use the EZmock\footnote{\url{https://github.com/cheng-zhao/EZmock}} method \cite{Chuang2015EZmock, Zhao2021}, which combines the computational efficiency of the Zeldovich approximation-based \citep{Zeldovich1970} gravitational evolution of the dark matter (DM) field with a flexible bias model and probability distribution function (PDF) mapping scheme that aims at fitting not only the reference two-point statistics on large scales, but also the bispectrum. Thanks to its speed and flexibility, EZmock was the baseline method to construct covariance matrices for the eBOSS Collaboration cosmological analyses \cite{Zhao2021}. In what follows we give a short description of the EZmock generation and processing to reproduce the observed DESI galaxy samples. For a full description of the mock suite, we refer the reader to ref.~\cite{KP3s8-Zhao}.

The EZmock suite consists of a set of 1000 North and an equal number of South Galactic Cap pseudo-lightcone\footnote{The $\Delta z$ between snapshots is large and the snapshots were joined for processing and addition of Fiber Assignment Effects. We do not use mocks with actual redshift evolution.} mocks for various tracers. For each tracer and redshift snapshot, 2000 $(6\Gpch)^3$ boxes of $1728^3$ DM particles were generated with independent Gaussian initial conditions. The resulting DM fields were converted to galaxy fields via the bias model, which is tuned snapshot-by-snapshot to the two-point clustering of the DESI Abacus HOD mocks (i.e. galaxies, see \cite{Yuan2024, antoine2023}), which were in turn fit to the data from the DESI one-percent survey \citep{DESI2023b.KP1.EDR}. These boxes were then converted to sky geometry. Note that, due to the large volume of the boxes, there is no volume replication required to fit one entire galactic cap. This is designed to avoid mis-estimating the cosmic variance. The different snapshots are then concatenated in a single pseudo-lightcone into which the DESI Y1 footprint and Fast Fibre Assignment (FFA) \cite{KP3s11-Sikandar} are applied to enhance the realism of the catalogues. A summary of the snapshots used for each of the DESI tracers can be found in \cref{tab:mocks}.

\begin{table}[t]
\centering
\caption{Redshift configuration of the EZmock suite for every DESI tracer used in this work. Note that most redshift bins are narrow enough so that only one snapshot redshift is a reasonable approximation for the whole range. Tuples denote $(z_{\rm min}, z_{\rm max})$.}
\label{tab:mocks}
\begin{tabular}{|l|ccc|cc|ccc|}
\hline
 & \multicolumn{3}{c|}{LRG} & \multicolumn{2}{c|}{ELG} & \multicolumn{3}{c|}{QSO} \\ \hline
 Snapshot $z$ & \multicolumn{1}{c|}{0.5} & \multicolumn{1}{c|}{0.8} & 1.1 & \multicolumn{1}{c|}{0.95} & 1.325 & \multicolumn{1}{c|}{1.1} & \multicolumn{1}{c|}{1.4} & 1.7 \\ \hline
Cut-sky $z$ & \multicolumn{1}{c|}{(0.4, 0.6)} & \multicolumn{1}{c|}{(0.6, 0.8)} & (0.8, 1.1) & \multicolumn{1}{c|}{(0.8, 1.1)} & (1.1, 1.6) & \multicolumn{3}{c|}{(0.8, 2.1)} \\ \hline
\end{tabular}
\end{table}

\subsection{Sample covariance estimation}


Given an ensemble of $N_{\rm mocks}$ measurements of a given $N_{\rm bins}$ bin-discretised statistic $\{S_i(x_j)\}_{i=0,j=0}^{N_{\rm mocks},N_{\rm bins}}$, the sample covariance is estimated as 
\begin{equation} \label{eq:sample-cov-def}
    \vb{C}_{{\rm s},ij} = \frac{1}{N_{\rm mocks} - 1}\sum_{k=1}^{N_{\rm mocks}} \left(S_k(x_i) - \bar{S}(x_i)\right)\left(S_k(x_j) - \bar{S}(x_j)\right),
\end{equation}
where, in our case, $S(x)$ is either the concatenated configuration space correlation function (2PCF) multipoles ($[\xi_{0}(s),\xi_{2}(s)]$) or power spectrum multipoles ($[P_{0}(k),P_{2}(k)]$).

It is well-known that a finite number of mocks used to build a covariance matrix induces a biased inverse covariance (also called precision) matrix and we apply the Hartlap correction factor \citep{Hartlap2007} to correct for this effect

\begin{equation}
    \tilde{\vb{C}}_{\rm m}^{-1} = \frac{N_{\rm mocks} - N_{\rm bins} - 2}{N_{\rm mocks} - 1}\vb{C}_{\rm s}^{-1}.
\end{equation}
These vary in value from 0.92 to 0.95 depending on the fitting setup.
The Percival correction \cite{Percival2014}
\begin{align}
    A &= \frac{2}{\left(N_{\rm mocks} - N_{\rm bins} - 1\right)\left(N_{\rm mocks} - N_{\rm bins} - 4\right)}\\
    B &= \frac{N_{\rm mocks} - N_{\rm bins} - 2}{\left(N_{\rm mocks} - N_{\rm bins} - 1\right)\left(N_{\rm mocks} - N_{\rm bins} - 4\right)}\\
    \vb{C}_{\rm m}^{-1} &= \frac{1 + A + B (N_{\rm par} + 1)}{1 + B(N_{\rm bins} - N_{\rm par})}\tilde{\vb{C}}_{\rm m}^{-1} \equiv \frac{1}{m_1}\tilde{\vb{C}}_{\rm m}^{-1},
\end{align}
is also applied to account for the biases in the estimated parameter covariance due to the estimated error in the sample covariance. Here, $N_{\rm par}$ is the number of free parameters used for the fit. This correction amounts to a factor of 1.02 to 1.05 depending on the setup. Tests performed in \cite{DESI2024.V.KP5} have shown that the combination of these two correction factors is consistent with the more recent formula provided in \cite{Percival2022}.

\subsection{Analytical covariance estimation}
(Semi-)Analytical covariance estimates are based on some input clustering measurements. Through this work, the analytical covariances have been tuned to the mock clustering (instead of data); therefore, we expect them to accurately match the sample covariance estimate. This design choice also makes our results applicable to clustering measurements other than \ezmock. Additionally, analytical covariances can use an estimate of the variance to tune the shot-noise contribution to a particular sample, thus partially emulating the non-Gaussianity. In this comparison, we test both an idealised case, in which the analytical covariances are using the mean variance of the complete mock clustering measurement sample, and the realistic case in which the covariances are generated from a single mock realisation and the noise is estimated using jackknife. Finally, the differences between data-based and mock-based covariances as well as differences between mock-generation techniques are not within the scope of this work. 

\subsubsection{Configuration Space: \rascalc{}}


The covariance matrix of the two-point correlation function of a single tracer can in general be written as
\begin{equation}
    \vb{C}(r,\mu,r',\mu') \equiv \mathrm{cov}\left[\xi(r,\mu), \xi(r',\mu')\right] = \langle\xi(r,\mu)\xi(r',\mu')\rangle - \langle\xi(r,\mu)\rangle\langle\xi(r',\mu')\rangle
\end{equation}
and involves contributions from quads of points, some of which may coincide, which are then related to the 4-, 3- and 2-point correlation functions.
This expression can be expanded into integrals that \rascalc{} \cite{rascal,rascal-jackknife,rascalC,rascalC-legendre-3} performs using importance sampling of groups of 2, 3 and 4-point groups from random catalogues generated for the survey volume; this ensures that the survey geometry is automatically taken into account in the integration.
A key feature of the \rascalc{} method is that instead of computing the exact non-Gaussian contributions, the Gaussian terms\footnote{Gaussian means based solely on the 2-point function. The 2-point function input to \rascalc{} is typically empirical and thus includes non-linearities.} are rescaled with a shot-noise rescaling parameter $\alpha_{\rm SN}$ in order to mimic the non-Gaussian contribution.
It is introduced through the modified shot-noise approximation (reducing the square of the overdensity appearing in the 3- and 2-point contributions):
\begin{equation}
    \langle\delta_i^2\rangle \approx \frac{\alpha_{\rm SN}}{n_i V_{\rm cell}}(1 + \delta_i),
\end{equation}
where $\alpha_{\rm SN}=1$ corresponds to the standard, Poissonian shot noise, $\delta_i$ and $n_i$ are the overdensity and the number density in cell $i$, and $V_{\rm cell}$ is the volume of this cell; the grid is chosen to be fine enough so that each cell contains at most 1 random point.
This approach has been shown to successfully mimic non-Gaussian effects in the covariance matrix, while the computations themselves remain Gaussian.
The shot-noise parameter can be tuned with a sample covariance (from mocks) or a jackknife covariance matrix directly from the data (we use both in this work).
The \rascalc{} method has already been successfully used in the DESI context for the BAO analysis of the early data\footnote{``Early DESI data'' designates the first two months of the main survey. It is part of the upcoming Data Release 1 \citep{DESI2024.I.DR1} and should not be confused with the Early Data Release \citep{DESI2023b.KP1.EDR}.} \cite{Moon2023,Misha2023}.
For further details on the features of the latest implementation of \rascalc{} we refer the reader to ref.~\cite{KP4s7-Rashkovetskyi}. 

\subsubsection{Fourier Space: \thecov}

Similarly to the configuration space case, the covariance of the power spectrum has contributions from up to the 4-point function or trispectrum. The Fourier space covariance, \thecov, is decomposed in a Gaussian $\vb{C}^{\rm G}$ and a non-Gaussian part $\vb{C}^{\rm T}$. However, in this work, we only use the Gaussian part, which can be written as

\def\PP{\mathcal{Q}_{\ell_1\ell_2}^{\ell_3\ell_4}(k_1,k_2)}
\def\PS{\mathcal{X}_{\ell_1\ell_2}^{\ell_3}(k_1,k_2)}
\def\SS{\mathcal{S}_{\ell_1\ell_2}(k_1,k_2)}

\begin{equation}
\begin{split}
    \vb{C}^{\rm G}_{\ell_1, \ell_2}(k_1, k_2) &= \sum_{\ell_3,\ell_4} \PP P_{\ell_3}(k_1) P_{\ell_4}(k_2)  \\
    &+ \sum_{\ell_3} \PS P_{\ell_3}(k_1) \\
    &+ \SS 
\end{split}
\end{equation}
where the kernels $\PP$, $\PS$ and $\SS$ respectively describe the cosmic variance (quadratic in $P_{\ell}$), variance-shot noise (linear in $P_{\ell}$) and shot noise contributions to the covariance. These kernels depend on the survey window and characteristics. For full details on the theory, including expressions for the kernels for cubic and sky geometries we refer the reader to ref.~\cite{Wadekar2019}. The implementation and tests on the inclusion of non-Gaussian terms in the context of DESI can be found in ref.~\cite{KP4s8-Alves}.

\subsection{Cosmological analyses}

The main aim of this paper is to provide information on how the construction of the covariance matrix affects cosmological results. To have results as close as possible to the real analysis, we use the combined clustering from the North and South galactic caps (NGC and SGC respectively). We do so for all tracers and redshift bins, which leads to 6 separate samples, 3 of which are LRGs, 2 ELGs and 1 QSOs (see \cref{tab:mocks} for $z$ ranges). This amounts to a total of about 6000 fits per cosmological analysis (see \cref{sec:bao-fit,sec:fs-fit}), some of which are very time- and computationally expensive. To minimise this burden, instead of running MCMC chains for every realisation we perform the minimisation using iMinuit \cite{Minuit}. It provides us with the necessary best-fit parameter estimates and a good approximation of their uncertainties. As a sanity check, we also run MCMC chains on the mean of 1000 mocks with the appropriate covariance scaling.


\subsubsection{BAO Fitting procedure}
\label{sec:bao-fit}
The increased precision of the Dark Energy Spectroscopic Instrument (DESI) motivates a revisit of the modelling and fitting procedure of the BAO signal. A comprehensive study of these techniques and their impact on the error budget of the experiment is presented in ref.~\cite{KP4s2-Chen}. We briefly summarise here the fitting methodology and highlight a few of the improvements over the legacy modelling strategy. 

BAO fits are performed on reconstructed configuration space measurements, i.e. the two first multipoles ($\ell=0,\,2$) of the two-point correlation function $\xi_\ell(s)$. BAO-reconstruction is a post-processing step that moves observed galaxies ``back in time'' using a (Zeldovich) approximated displacement field, estimated using the Iterative Fast Fourier Transform (IFFT) method first presented in ref.~\cite{Burden2015} and optimised for DESI in \cite{KP4s3-Chen, KP4s4-Paillas}. The systematic tests and hyperparameter choices for the method are described in ref.~\cite{KP4s3-Chen,KP4s4-Paillas}. The eBOSS analyses used a ``RecIso'' reconstruction convention, where the galaxies are moved back and RSD is approximately removed, whereas for the random sample no RSD-removal term is used. As a result, the final clustering will show little to no RSD (quadrupole) signal. DESI 2024 analyses, on the other hand, use the ``RecSym'' reconstruction convention \cite{White2015} -- which stands for symmetric reconstruction -- where both data and randoms are treated in the same way, that is, they are displaced with the same field (including the RSD term), which preserves the quadrupole signal. 

We measure the dilations of the BAO scale in the parallel ($\alpha_\parallel$) and perpendicular ($\alpha_\perp$) directions, with respect to a template or fiducial cosmology. This means the relevant BAO parameters are interpreted as
\begin{align}
    \alpha_\parallel &= \frac{H^{\rm fid}(z)}{H(z)}\frac{r_d^{\rm fid}}{r_d},\\
    \alpha_\perp &= \frac{D_A(z)}{D_A^{\rm fid}(z)}\frac{r_d^{\rm fid}}{r_d},
\end{align}
where $H^{(\rm fid)}(z)$ denotes the Hubble parameter, $D_A^{(\rm fid)}(z)$ the angular diameter distance, $r_d$ the sound horizon at the drag epoch and the superscript ${(\rm fid)}$ denotes quantities computed assuming a fiducial cosmology.

In practice, the template is modified by the scaling of the (observed) Fourier space coordinates $(k,\mu)$ as
\begin{align}
    k' &= \frac{k}{\alpha_\perp}\left[1 + \mu^2\left(\frac{\alpha_\parallel^2}{\alpha_\perp^2} - 1\right)\right]^{1/2} \\ 
    \mu' &= \mu\frac{\alpha_\perp}{\alpha_\parallel}\left[1 + \mu^2\left(\frac{\alpha_\parallel^2}{\alpha_\perp^2} - 1\right)\right]^{1/2}.
\end{align}

One of the major changes in the revised BAO modelling for DESI is that, even for a configuration space, the $\alpha$ scaling is performed directly in the Fourier-space template, instead of being applied to the model correlation function -- as was previously the case -- (see equation 2 in ref.~\cite{Xu2012}). While the relevant dilation parameters are $\alpha_{\perp,\parallel}$, we will more often show results in terms of the strictly anisotropic $\alpha_{\rm ap} \equiv \alpha_\parallel/\alpha_\perp$ and isotropic $\alpha_{\rm iso}\equiv \alpha_\parallel^{1/3}\alpha_\perp^{2/3}$ dilations.

Similarly to the legacy approach, the template relies on the separation of the wiggle ($P_{\rm w}$) and non-wiggle ($P_{\rm nw}$) component of the power spectrum. In a general form, this is written as
\begin{align}
    P_{gg,\ell}(k) &= \frac{2\ell + 1}{2}\int_{-1}^1\dd\mu \left[\mathcal{B}(k,\mu)P_{\rm nw}(k) + \mathcal{C}(k,\mu)P_{\rm w}(k)\right]L_\ell(\mu)\\
    \xi_{gg,\ell}(s) &= i^\ell\int\frac{\dd k}{2\pi^2} k^2 P_{gg,\ell}(k)j_\ell(ks) + \tilde{\mathcal{D}}_\ell(s)
\end{align}

For 2D BAO fits we adopt the RecSym reconstruction convention (see refs.~\cite{KP4s3-Chen, KP4s4-Paillas} for details on reconstruction algorithms and conventions), such that

\begin{align}
\mathcal{B}(k, \mu) & = \left(b + f\mu^2\right)^2\left[1 + \frac{1}{2}k^2\mu^2\Sigma_s^2\right]^{-2},\\
\mathcal{C}(k, \mu) & = \left(b + f\mu^2\right)^2\exp{-\frac{1}{2}k^2\left[\mu^2\Sigma_\parallel^2 + (1-\mu)^2\Sigma_\perp^2\right]},\\
\tilde{\mathcal{D}}_0(s) &= \tilde{a}_{0,0} + \tilde{a}_{0,1}\left(\frac{sk_{\rm min}}{2\pi}\right)^2,\\
\tilde{\mathcal{D}}_2(s) &= \tilde{a}_{2,0} + \tilde{a}_{2,1}\left(\frac{sk_{\rm min}}{2\pi}\right)^2 + \Delta^3\left[a_{2,0}B_{2,0}(s\Delta)+a_{2,1}B_{2,1}(s\Delta)\right].
\end{align}

This parameterisation ensures that there is no broadband component ($\tilde{\mathcal{D}}$) capable of modelling the BAO feature on its own given that it is parameterised by a spline basis separated by $\Delta = 0.06\hMpc > 2\pi / s_{\rm BAO}$ where $s_{\rm BAO}$ is the BAO scale. The explicit expressions for the $B_{i,j}$ functions can be found in appendix E of ref.~\cite{KP4s2-Chen}. The resulting BAO model, similarly to previous approaches, has the dilation factors $\alpha_{\perp,\parallel}$, bias $b$, the growth rate of structure $f$, the parallel and perpendicular BAO damping $\Sigma_{\parallel,\perp}$ and the finger of God $\Sigma_s$ parameters in addition to the nuisance broadband parameters $a_{i,j},\,\tilde{a}_{i,j}$. For BAO fits, we restrict the fitting range to $[s_{\rm min},s_{\rm min}] = [50, 150]\Mpch$ using bins of $\Delta s = 4\Mpch$, the default choice in the DESI 2024 cosmological analysis and $k_{\rm min} = 0.02 \hMpc$. Further information, such as prior ranges, can be found in ref.~\cite{KP4s2-Chen}.

\subsubsection{Full-Shape analysis}
\label{sec:fs-fit}

To extract information about the growth of structure, Full-Shape analyses, rely on perturbation theory (PT) codes such as \texttt{velocileptors} \cite{Chen2020,Chen2021,KP5s2-Maus} or \texttt{FOLPS} \cite{Noriega2022,KP5s3-Noriega} that predict the redshift-space galaxy power spectrum $P_{gg}(\vb{k})$ given a linear matter power spectrum $P_{\rm lin}(k)$ as provided by a Boltzmann code (e.g. \texttt{CLASS} \citep{CLASS} or \texttt{CAMB} \citep{CAMB,Howlett2012}). In this work, we use the \texttt{velocileptors} code in order to predict the redshift-space galaxy power spectrum as 
\begin{multline}
    P_{gg}(\vb{k}) = P_{gg}^{\rm PT}(\vb{k}) + (b+f\mu^2)(b\alpha_0 + f \alpha_2\mu^2+f^2\alpha_4\mu^4)k^2P_{\rm s,b1^2}(\vb{k}) \\
    + (\mathrm{SN}_0 + \mathrm{SN}_2k^2\mu^2 + \mathrm{SN}_4k^4\mu^4),
\end{multline}
where $b$ and $f$ are still the linear bias and growth rate parameters, whereas $\alpha_i$ and $\mathrm{SN}_j$ are nuisance parameters. Some of these nuisance parameters can be theoretically correlated, thus reducing the freedom of the model but, in turn, making the fits easier. In this work we default to the use of ``maximal freedom'' settings where the nuisance parameters are independent from each other. A thorough description of the perturbation theory term $P_{gg}^{\rm PT}(\vb{k})$ as well as the extra terms is provided in \cite{KP5s1-Maus}. Given the theoretical prediction, the observed power spectrum is obtained through
\begin{equation}
    P_{gg}^{\rm obs}(\vb{k}_{\rm obs}) = \alpha_\perp^{-2}\alpha_\parallel^{-1}P_{gg}(\vb{k}^{\rm true})
\end{equation}
where $\alpha_\perp$, $\alpha_\parallel$ are the BAO dilation factors. 

For these Full-Shape fits, we use power spectra measured in the range $k \in [0.02, 0.2]\hMpc$ in bins of $\Delta k = 0.005\hMpc$ and the baseline settings of \cite{DESI2024.V.KP5}. Furthermore, we apply both window and wide-angle corrections to the theory power spectra in order to properly account for survey effects on the fit. For details on how these are estimated, we refer the reader to refs.~\cite{DESI2024.II.KP3,Beutler2021} and references therein.

\paragraph{ShapeFit:}
Similar to other parameter compression techniques (such as the BAO analysis), ShapeFit \cite{Brieden21} relies on a single template linear power spectrum evaluated at a reference (or template) cosmology. This implies that it also constrains dilation parameters in addition to the growth rate parameter $f\sigma_8$ and a new parameter $m$ introduced to account for the cosmology-dependent shape variation around the default linear power spectrum, which itself depends on $\omega_m$ and $\omega_b$. The template linear power spectrum is therefore given as
\begin{equation}
    P'_{\rm lin}(\vb{k}) = P_{\rm lin}(\vb{k})\exp{\frac{m}{a}\tanh(a\ln(\frac{k}{k_p})) + n\ln(\frac{k}{k_p})},
\end{equation}
where $k_p \approx 0.03\hMpc$ is the pivot wavenumber where the baryon suppression slope reaches its maximum, the parameter $n$ controls the scale-independent slope of the power spectrum and is set to 0 for all fits here. The amplitude parameter $a$ is fixed to $a = 0.6$. Our results are shown in terms of parameters relative to a reference, that is, we report $dm\equiv m-1$ instead of $m$ and $df\equiv f/f^{\rm fid}$ instead of $f$.

The main advantage of this template method is precisely that it is possible to reinterpret the compressed parameters in terms of various cosmological models without needing to rerun the potentially expensive fits. Moreover, these fits are not too expensive when using template methods as there is no call to a Boltzmann code for every likelihood evaluation compared to the direct fitting technique. Nonetheless, the perturbation theory integrals required to predict the galaxy power spectrum remain moderately expensive and we use a Taylor-expansion emulator\footnote{\url{https://github.com/cosmodesi/desilike.git}} to bypass this cost.

\paragraph{Direct fit:}

An alternative to template methods is of course modelling the galaxy power directly by providing a linear power spectrum evaluated at each cosmology, corresponding to a likelihood evaluation (instead of modifying a template from a fiducial cosmology). As mentioned before, this is much more expensive computationally but allows to directly constrain the cosmological parameters of a given model. In the case of $\Lambda$CDM, for example, using galaxy clustering measurements, it is possible to constrain $\omega_{\rm cdm}, H_0$ and $\log(10^{10}A_s)$ with the help of some (well informed) priors on $n_s, M_\nu$ and $\omega_{\rm b}$, which are not part of the baseline Full-Shape analysis. In this case, the computational cost of the extra Boltzmann call per likelihood estimation can also be offset by an emulator, which we do in the present work. A full description of the prior ranges for all the cosmological and nuisance parameters is provided in ref.~\cite{KP5s1-Maus}.

\section{Results}
\label{sec:results}
\subsection{Post-reconstruction configuration space covariance and BAO}
\label{sec:conf-results}
As stated before, we focus on the BAO analysis in configuration space and show only RecSym-reconstructed samples and non-Gaussian (rescaled) \rascalc{} covariances unless explicitly stated otherwise. 
\begin{figure}
    \centering
    \includegraphics[width=\textwidth]{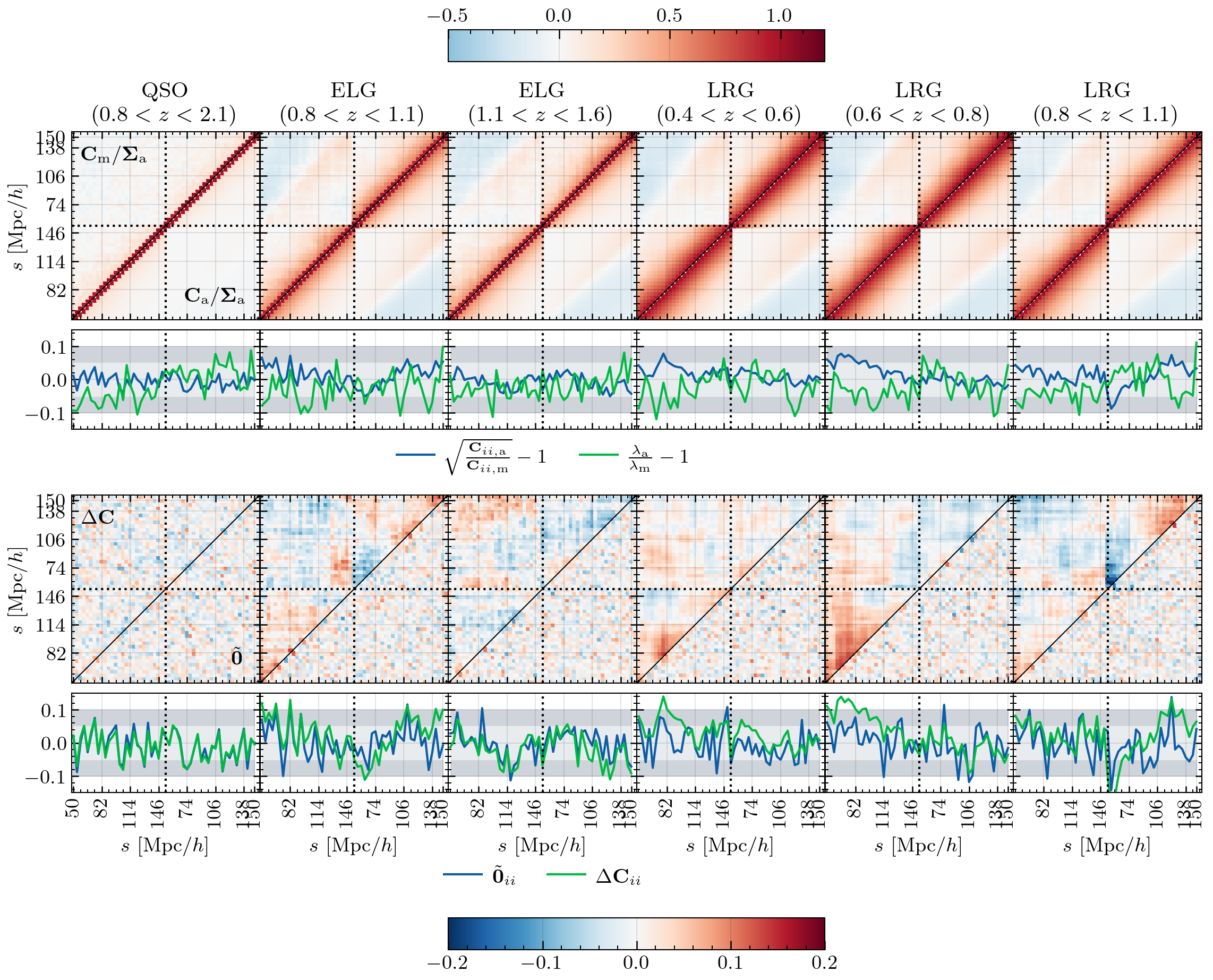}
    \caption{Visual comparison of analytical and mock covariances in configuration space for the $\ell=0,\,2$ correlation function multipoles. The \rascalc{} matrix has been tuned to the ensemble of mocks. \textit{Top row:} The top panels show the covariance matrices normalised by the variance of the analytical matrix ($\vb{\Sigma}_{ij,\rm a}\equiv \sqrt{\vb{C}_{ii,\rm a}\vb{C}_{jj,\rm a}}$, so both diagonals are not 1). The upper triangle corresponds to the mock $\vb{C}_{\rm m}$ and the bottom to the analytical $\vb{C}_{\rm a}$ covariance.  The subpanels in the top row show the fractional difference of standard deviations and sorted eigenvalues $\lambda$. \textit{Bottom row:} The top triangles in the bottom row show $\Delta\vb{C}\equiv\frac{\vb{C}_{\rm a} - \vb{C}_{\rm m}}{\Sigma_{\rm a}}$ the normalised difference of covariances. Similarly, the bottom triangle shows the symmetric inverse test $\vb{\tilde{0}} = \vb{I} - \vb{C}_{\rm a}^{-1/2}\vb{C}_{\rm m}\vb{C}_{\rm a}^{-1/2}$. Overall the matrices on the bottom should equal zero everywhere. The subpanels in the bottom row show the diagonals of the matrices above. The shaded regions in the subpanels highlight 5 and 10\% regions.}
    \label{fig:visual-conf}
\end{figure}

As a first test, we compare the covariance matrices from the \ezmock, $\vb{C}_{\rm m}$, and \rascalc{}, $\vb{C}_{\rm a}$ in \cref{fig:visual-conf}. The top row shows both matrices normalised by the variance of the analytical one, $\vb{\Sigma}_{\rm a}$ for all tracers. We purposefully use the same normalisation for both matrices to highlight any possible differences in the diagonals. The corresponding sub-panels show the fractional difference in standard deviations from both estimates $\sqrt{\vb{C}_{{ii}, \mathrm{a}} / \vb{C}_{ii,\rm m}} - 1$. For the monopole, the analytical standard deviation is $\sim 5\%$ larger than the mock estimate, especially in the lower redshift bins and towards the low end of the fitting range. On the other hand, the quadrupoles seem consistent within 5\% for most tracers and redshift bins, with the exception of the $0.8 < z < 1.1$ range where both LRGs and ELGs show a $-5$ to $-10\%$ discrepancy towards low $s$, which turns positive towards the high limit of the fitting range.

Moreover, the same sub-panel shows the eigenvalue ratios $\lambda_{\rm a} / \lambda_{\rm m} - 1$. These however do not correspond to the abscissas as they are sorted by default. We observe consistently $\sim 5 - 10\%$ smaller analytical eigenvalues for the largest 20 or so values.
 
The bottom row of \cref{fig:visual-conf} shows an element-by-element direct comparison of the covariance matrices. The upper triangle of the top panel shows the simple normalised difference $\Delta \vb{C} = (\vb{C}_{\rm a} - \vb{C}_{\rm m}) / \vb{\Sigma}_{\rm a}$. For the LRG samples, there are differences of around 15\%, especially in the mono-mono quadrant of the matrix. Additionally, the $0.8<z<1.1$ matrices show similar trends in the quadrupole-quadrupole quadrant, also with over 10\% differences. This is highlighted in the respective sub-panels. It is clear, however, that as redshift increases, $\Delta\vb{C}$ approaches a more uniform zero matrix. Finally, the lower triangles of these same panels show the $\vb{\tilde{0}} = \vb{I} - \vb{C}_{\rm a}^{-1/2}\vb{C}_{\rm m}\vb{C}_{\rm a}^{-1/2}$ matrix as for of an inverse-test. This matrix should be zero if the analytical and covariance matrices are similar enough so that $\vb{C}_{\rm a}^{-1}\vb{C}_{\rm m}\approx \vb{I}$, the identity matrix. This inverse-test statistics are uniform and consistent with zero within 0.05, pointing at a good agreement between the matrices.
 




\begin{figure}
    \centering
    \includegraphics[width=\textwidth]{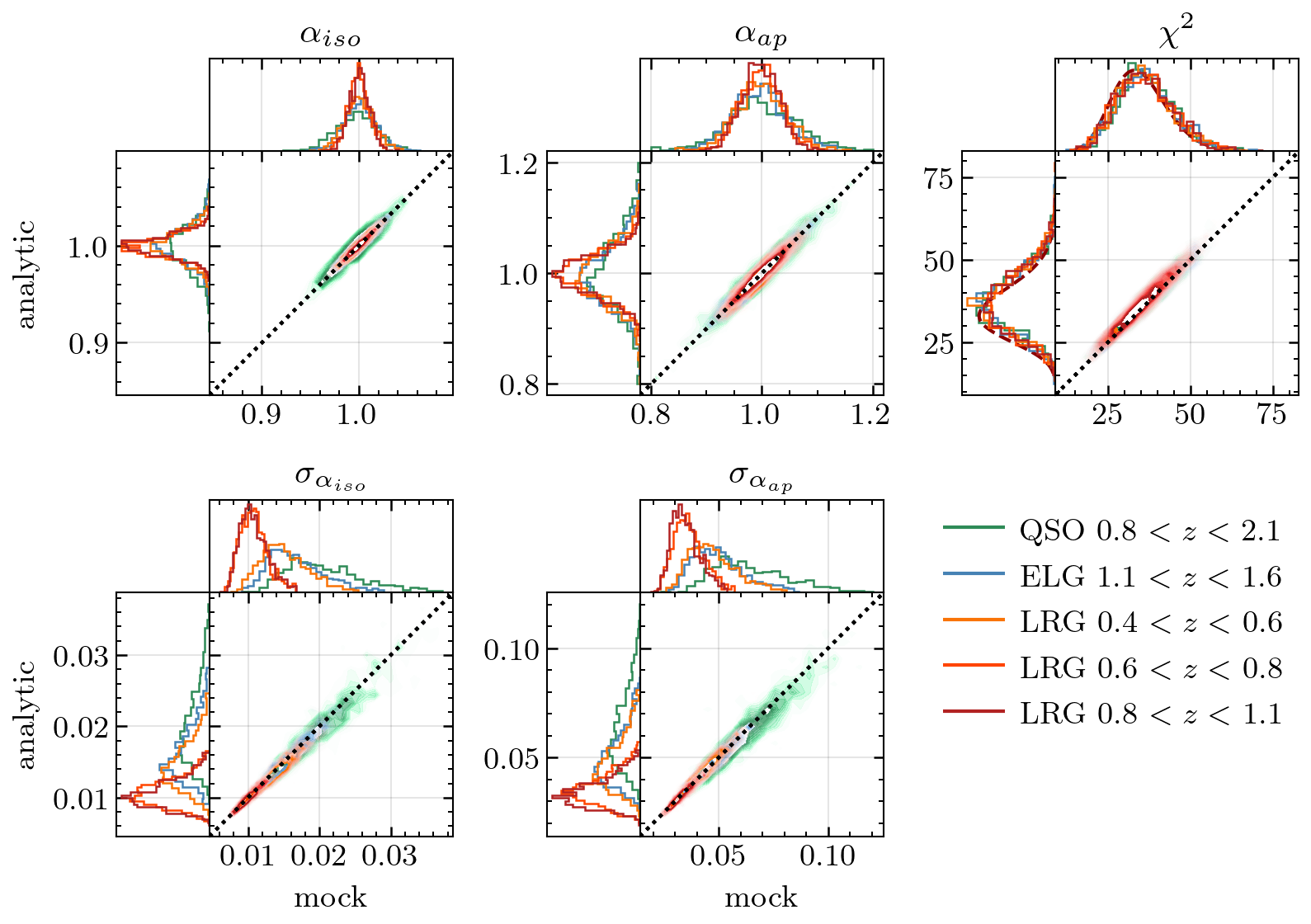}
    \caption{2D BAO fit comparison for the observed samples. The top row shows the comparison plots for the relevant dilation parameters as well as the $\chi^2$ goodness-of-fit comparison. The bottom row shows the comparison of the error estimates for the dilation parameters.}
    \label{fig:avg-lrg-hiz}
\end{figure}

\subsubsection{BAO analysis}
In this subsection, we evaluate the effect of the differences in the covariances in terms of the BAO parameters and their error estimates, as well as the goodness-of-fit. \Cref{fig:avg-lrg-hiz} shows the results of fitting the ensemble of mocks for a given tracer and redshift bin. We fit each mock both with the analytical covariance (in this case tuned to the ensemble of mocks) and with a sample covariance built with the other 999 realisations. For each panel, we compute a few relevant statistics, such as the mean percent difference and the correlation coefficient $r$.

The summary statistics obtained for each individual sample are shown on \cref{tab:bao-avg-all-tracers}.

We do not observe any significant difference $(<0.07\%)$ in terms of parameter estimates and the ensemble of measurements follows a Gaussian distribution with width approximately the mean of the uncertainties. On the other hand, the errors on the parameters show slight discrepancies where the analytic covariance underestimates the $\sigma_{\alpha_{iso}}$ and $\sigma_{\alpha_{ap}}$ by around 3\%. In terms of individual samples, the largest discrepancy is $<4\%$. We suspect this larger difference is compatible with a statistical fluke. Accordingly, the $\chi^2$ values are $2.8\pm0.3\%$ higher for the analytic covariance than they are for the sample estimate which is small enough for us to consider the matrix estimates equivalent. In the side panels of the top-right panel, we have also added the theoretical $\chi^2$ distribution with the appropriate number of degrees of freedom as a dashed line. Notice that neither covariance results in a $\chi^2$ distribution that follows the theoretical expectation, in the case of the sample covariance, this is somewhat expected due to the limited number of mocks.

Additionally, we evaluate that the Pearson correlation coefficient $r$ is on average $0.974\pm0.005$ for $\alpha_{iso}$ and $0.927\pm0.018$ for $\alpha_{ap}$. Similarly, the corresponding error estimates show a correlation of $0.9726\pm0.006$ and $0.931\pm0.014$. Overall, the samples are almost perfectly correlated ($r\approx1$), which also supports the conclusion that the different matrix estimates do not affect cosmological results significantly.

\begin{table}[t]
\centering
\caption{Summary of statistics comparing the performance of covariances in BAO fits for all tracers and $z$ bins. The mean percent difference $\Delta(X)\equiv100\times\langle \frac{X_{\rm m}}{X_{\rm a}} - 1\rangle$ and $r(X$ denotes the correlation coefficient between the samples of statistic $X$.}
\label{tab:bao-avg-all-tracers}
\rowcolors{1}{white}{mygray}
\begin{tabular}{ccccccc}
\hline
                                   &  \begin{tabular}[c]{@{}c@{}}QSO\\$(0.8, 2.1)$\end{tabular}  &  \begin{tabular}[c]{@{}c@{}}ELG\\$(1.1, 1.6)$\end{tabular}  &  \begin{tabular}[c]{@{}c@{}}LRG\\$(0.4, 0.6)$\end{tabular}  &  \begin{tabular}[c]{@{}c@{}}LRG\\$(0.6, 0.8)$\end{tabular}  &  \begin{tabular}[c]{@{}c@{}}LRG\\$(0.8, 1.1)$\end{tabular}  &  Avg.  \\
\hline
     $\Delta({\alpha_{iso}})$      &                            -0.07                            &                            0.02                             &                            -0.01                            &                            0.01                             &                            -0.01                            & -0.01  \\
        $r({\alpha_{iso}})$        &                            0.95                             &                            0.98                             &                            0.97                             &                            0.98                             &                            0.98                             &  0.97  \\
 $\Delta({\sigma_{\alpha_{iso}}})$ &                            2.79                             &                            2.06                             &                            3.80                             &                            -0.64                            &                            1.85                             &  1.97  \\
   $r({\sigma_{\alpha_{iso}}})$    &                            0.90                             &                            0.93                             &                            0.93                             &                            0.94                             &                            0.95                             &  0.93  \\
      $\Delta({\alpha_{ap}})$      &                            0.02                             &                            0.01                             &                            0.10                             &                            -0.07                            &                            0.00                             &  0.01  \\
        $r({\alpha_{ap}})$         &                            0.95                             &                            0.98                             &                            0.96                             &                            0.98                             &                            0.98                             &  0.97  \\
 $\Delta({\sigma_{\alpha_{ap}}})$  &                            2.91                             &                            0.19                             &                            -0.15                            &                            1.43                             &                            1.37                             &  1.15  \\
    $r({\sigma_{\alpha_{ap}}})$    &                            0.91                             &                            0.94                             &                            0.93                             &                            0.95                             &                            0.95                             &  0.94  \\
        $\Delta({\chi^2})$         &                            -2.57                            &                            -2.40                            &                            -3.10                            &                            -2.89                            &                            -2.96                            & -2.78  \\
           $r({\chi^2})$           &                            0.98                             &                            0.98                             &                            0.98                             &                            0.98                             &                            0.98                             &  0.98  \\
\hline
\end{tabular}
\end{table}

While satisfactory, it must be noted that the analytic covariance so far was tuned to a full sample of mocks, that is, the input clustering was the average of 1000 mock realisations (making it very smooth) and the \rascalc{} shot noise rescaling parameter was tuned using the sample variance. This is far from the realistic case in which the covariance is estimated from the data (i.e. a single, noisy realisation) and the rescaling is estimated from a smaller sample of jackknife measurements. To validate these single-realisation covariances, we created 10 different analytical matrices based on 10 of the \ezmock following the procedure used fot data-based covariances and repeated the test in \cref{fig:avg-lrg-hiz} for the LRG3 sample. \Cref{tab:bao-ind-lrg} shows the mean and error on the mean of the statistics $\Delta(X)$ and $r(X)$ over the 10 \rascalc{} covariances. We observe that both the percent differences and correlation coefficients obtained for the same sample with the \rascalc{} matrix generated from the mock ensemble are mostly consistent with the errors estimated from the 10 covariances, with the larger discrepancies being in the estimation of the errors $\sigma_X$.  We conclude then that the use of independent mocks to produce the analytical covariance does not bias either the parameter estimates or their errors and has little effect on the goodness of fit. While a larger covariance sample would be ideal to further test these observations, this remains prohibitive due to the computational cost of computing a single \rascalc{} covariance.

\begin{table}[t]
\centering
\caption{Summary of the statistics derived from fitting the mock ensemble with 10 different analytical covariance matrices. We report the mean and error ($\sigma / \sqrt{10}$) of the $\Delta(X)$ and $r(X)$ over the 10 different \rascalc{} covariances.
Same conventions as \cref{tab:bao-avg-all-tracers}.}
\label{tab:bao-ind-lrg}
\rowcolors{1}{white}{mygray}
\begin{tabular}{cl}
\hline
                                  & LRG (0.8 < $z$ < 1.1)                    \\
\hline
     $\Delta(\alpha_{iso})$      & $-0.00923 \pm 0.00010$ \\
        $r(\alpha_{iso})$        & $0.97933 \pm 0.00004$  \\
 $\Delta(\sigma_{\alpha_{iso}})$ & $0.2 \pm 0.16$          \\
   $r(\sigma_{\alpha_{iso}})$    & $0.9536 \pm 0.0009$    \\
      $\Delta(\alpha_{ap})$      & $-0.0024 \pm 0.0005$   \\
        $r(\alpha_{ap})$         & $0.98027 \pm 0.0006$  \\
 $\Delta(\sigma_{\alpha_{ap}})$  & $-0.2 \pm 0.16$         \\
    $r(\sigma_{\alpha_{ap}})$    & $0.9471 \pm 0.0006$    \\
        $\Delta(\chi^2)$         & $1.8 \pm 0.47$          \\
           $r(\chi^2)$           & $0.98079 \pm 0.00002$  \\
\hline
\end{tabular}
\end{table}

\subsection{Fourier space and the full shape analysis}
\label{sec:fourier-results}
\begin{figure}
    \centering
    \includegraphics[width=\textwidth]{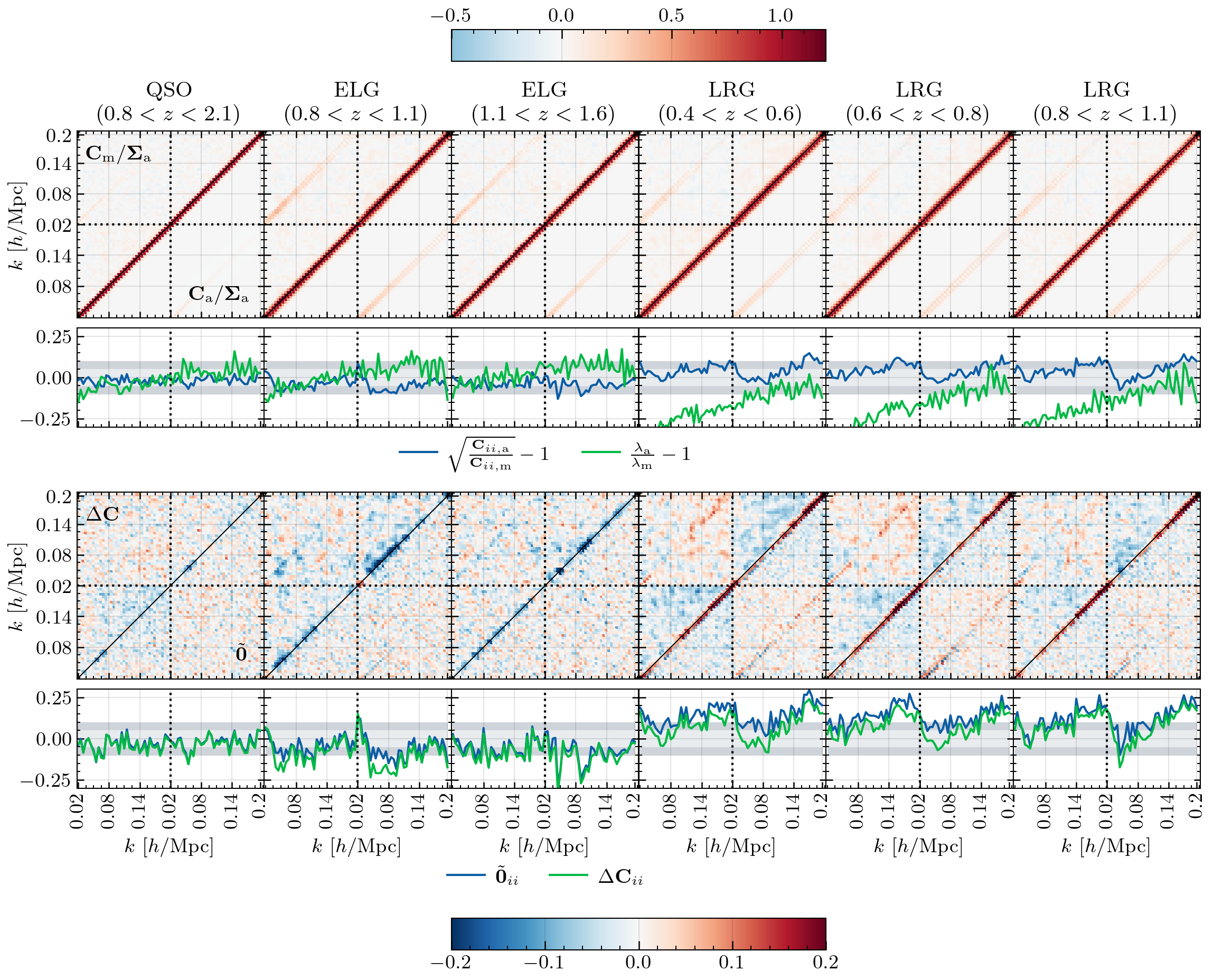}
    \caption{Visual comparison of analytical and mock covariances in Fourier space for the $\ell=0,\,2$ power spectrum multipoles. The \thecov matrix has been tuned to the ensemble of mocks. Same convention as \cref{fig:visual-conf}.}
    \label{fig:visual-fourier}
\end{figure}


\Cref{fig:visual-fourier} shows tests equivalent to those in \cref{fig:visual-conf} in Fourier space. From the top row, it is evident that the mock covariance has many nonzero off-diagonal elements which the analytical covariance does not estimate in the Gaussian approximation, this hints at the necessity of the inclusion of the trispectrum terms which yield nonzero off-diagonal contributions. The ratio of standard deviations shown in the subpanels on the top row shows that the shot noise rescaling that was implemented can reproduce the variance in the mock covariances to a $5\%$ level on all scales, however, the same panel shows that the eigenvalues are not properly reproduced, especially for lower redshift, LRG samples. The bottom row shows the difference and inverse tests which show a $\sim 10\%$ to $20\%$ discrepancy for the LRG samples which increases $\propto k$. For ELG and QSO samples there is little scale dependence and discrepancies between estimates lie around the $5\%$ mark. In the Fourier space, the $\vb{\tilde{0}}$ matrix shows non zero diagonal elements in the monopole-quadrupole quadrant.

Preliminary tests with mocks without FFA and the full DESI 5-year footprint showed that there was no significant difference between the mock and analytic estimates in terms of cosmological parameters. Changes in any of these aspects in principle could cause the discrepancies observed for the covariance matrices themselves.

\subsubsection{ShapeFit}

\begin{figure}
    \centering
    \includegraphics[width=\textwidth]{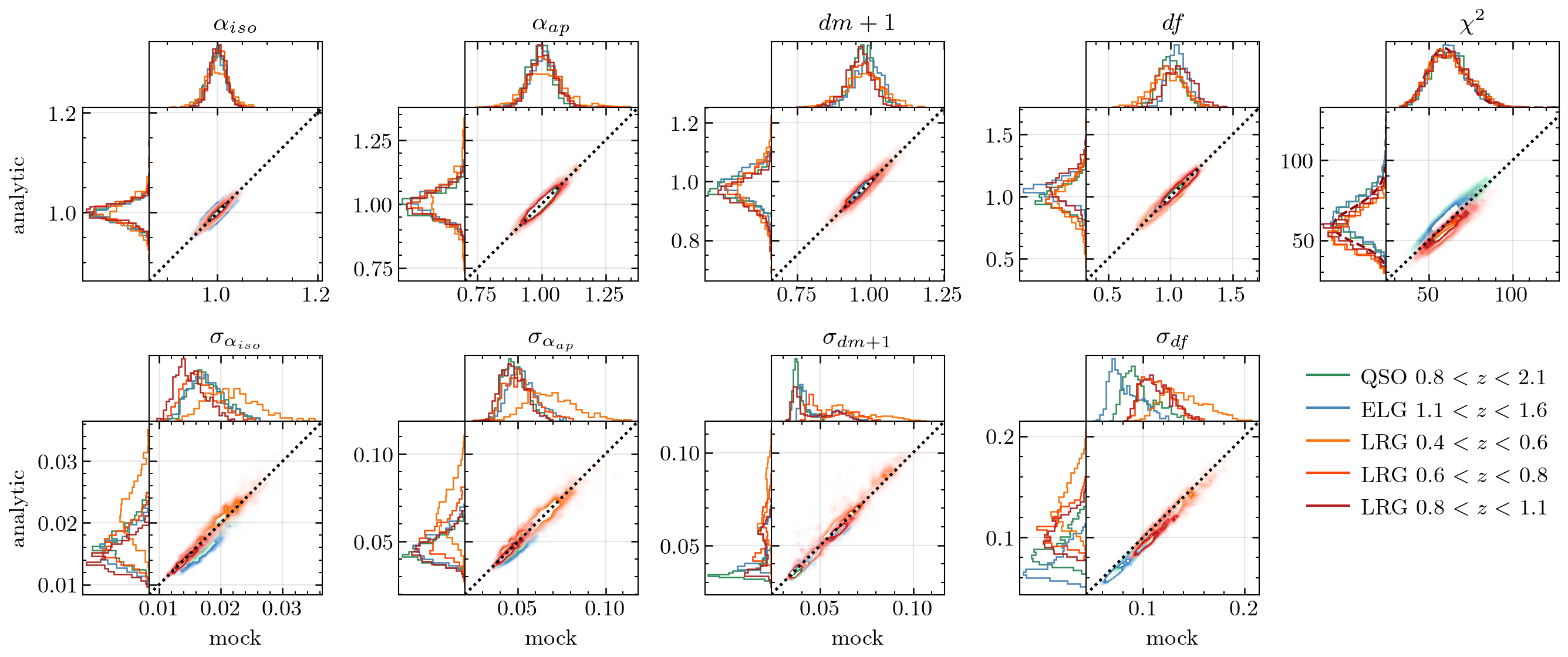}
    \caption{ShapeFit comparison for the observed samples. The top row shows the comparison plots for the relevant parameters as well as the $\chi^2$ goodness-of-fit comparison. The bottom row shows the comparison of the error estimates for the parameters.}
    \label{fig:shapefit-avg-lrg-lowz}
\end{figure}

\Cref{fig:shapefit-avg-lrg-lowz} shows the differences in the ShapeFit parameters and their respective error estimates. Similarly to the BAO analysis, the parameter estimates show on average no significant bias due to the covariance matrix choice. However, the error estimates are significantly affected by the matrix choice, with biases of around 3\% for the dilation parameters, 3\% for $m$ and 9\% for $f$. The individual measurements for the comparison statistics can be found in \cref{tab:shapefit-avg-all-tracers}. From these tests, we conclude that the analytical covariance estimate in Fourier space does not provide results consistent with the corresponding sample covariance in terms of ShapeFit parameters.

\begin{table}
    \centering
    \rowcolors{1}{white}{mygray}
\begin{tabular}{ccccccc}
\hline
                                   &  \begin{tabular}[c]{@{}c@{}}QSO\\$(0.8, 2.1)$\end{tabular}  &  \begin{tabular}[c]{@{}c@{}}ELG\\$(1.1, 1.6)$\end{tabular}  &  \begin{tabular}[c]{@{}c@{}}LRG\\$(0.4, 0.6)$\end{tabular}  &  \begin{tabular}[c]{@{}c@{}}LRG\\$(0.6, 0.8)$\end{tabular}  &  \begin{tabular}[c]{@{}c@{}}LRG\\$(0.8, 1.1)$\end{tabular}  &  Avg.  \\
\hline
     $\Delta({\alpha_{iso}})$      &                            0.02                             &                            0.04                             &                            -0.04                            &                            -0.04                            &                            -0.02                            & -0.01  \\
        $r({\alpha_{iso}})$        &                            0.97                             &                            0.90                             &                            0.89                             &                            0.88                             &                            0.96                             &  0.92  \\
 $\Delta({\sigma_{\alpha_{iso}}})$ &                            4.34                             &                            11.75                            &                            -0.75                            &                            0.22                             &                            2.21                             &  3.55  \\
   $r({\sigma_{\alpha_{iso}}})$    &                            0.94                             &                            0.92                             &                            0.84                             &                            0.92                             &                            0.95                             &  0.91  \\
      $\Delta({\alpha_{ap}})$      &                            -0.16                            &                            0.23                             &                            0.20                             &                            0.35                             &                            0.19                             &  0.16  \\
        $r({\alpha_{ap}})$         &                            0.96                             &                            0.89                             &                            0.87                             &                            0.85                             &                            0.95                             &  0.90  \\
 $\Delta({\sigma_{\alpha_{ap}}})$  &                            6.36                             &                            10.03                            &                            2.35                             &                            -2.62                            &                            1.65                             &  3.55  \\
    $r({\sigma_{\alpha_{ap}}})$    &                            0.93                             &                            0.87                             &                            0.77                             &                            0.85                             &                            0.92                             &  0.87  \\
         $\Delta({dm+1})$          &                            -0.06                            &                            -0.06                            &                            0.26                             &                            0.09                             &                            -0.06                            &  0.03  \\
            $r({dm+1})$            &                            0.97                             &                            0.95                             &                            0.92                             &                            0.93                             &                            0.96                             &  0.95  \\
     $\Delta({\sigma_{dm+1}})$     &                            8.08                             &                            6.32                             &                            -1.93                            &                            0.41                             &                            0.41                             &  2.66  \\
       $r({\sigma_{dm+1}})$        &                            0.84                             &                            0.93                             &                            0.84                             &                            0.85                             &                            0.88                             &  0.87  \\
          $\Delta({df})$           &                            0.08                             &                            -0.13                            &                            0.19                             &                            -0.21                            &                            -0.05                            & -0.02  \\
             $r({df})$             &                            0.97                             &                            0.96                             &                            0.92                             &                            0.92                             &                            0.96                             &  0.95  \\
      $\Delta({\sigma_{df}})$      &                            8.16                             &                            10.88                            &                            4.18                             &                            2.43                             &                            8.78                             &  6.89  \\
        $r({\sigma_{df}})$         &                            0.92                             &                            0.93                             &                            0.85                             &                            0.88                             &                            0.92                             &  0.90  \\
        $\Delta({\chi^2})$         &                            -3.46                            &                            -5.18                            &                            12.56                            &                            10.13                            &                            6.09                             &  4.03  \\
           $r({\chi^2})$           &                            0.97                             &                            0.97                             &                            0.96                             &                            0.96                             &                            0.96                             &  0.96  \\
\hline
\end{tabular}

    \caption{Comparison statistics for ShapeFit results in Fourier space with different samples. Same convention as \cref{tab:bao-avg-all-tracers}.}
    \label{tab:shapefit-avg-all-tracers}
\end{table}
To rule out problems with the fitting pipeline and in particular in our choice of minimiser, we perform equivalent tests in configuration space using the corresponding pre-reconstruction RascalC covariance matrices. These fits are performed in the range of $s\in[30,150]\hMpc$ with a binning of $4\hMpc$. These results are listed in \cref{tab:shapefit-avg-all-tracers-rascalc}. Overall, the agreement between the analytic and mock covariances in the configuration space, in terms of error estimates, is better than in Fourier space by as much as a factor 2 (e.g. $\sigma_{df}$). 

\begin{table}
    \centering
    \rowcolors{1}{white}{mygray}
\begin{tabular}{ccccccc}
\hline
                                   &  \begin{tabular}[c]{@{}c@{}}QSO\\$(0.8, 2.1)$\end{tabular}  &  \begin{tabular}[c]{@{}c@{}}ELG\\$(1.1, 1.6)$\end{tabular}  &  \begin{tabular}[c]{@{}c@{}}LRG\\$(0.4, 0.6)$\end{tabular}  &  \begin{tabular}[c]{@{}c@{}}LRG\\$(0.6, 0.8)$\end{tabular}  &  \begin{tabular}[c]{@{}c@{}}LRG\\$(0.8, 1.1)$\end{tabular}  &  Avg.  \\
\hline
     $\Delta({\alpha_{iso}})$      &                            -0.03                            &                            -0.05                            &                            -0.01                            &                            -0.04                            &                            -0.03                            & -0.03  \\
        $r({\alpha_{iso}})$        &                            0.94                             &                            0.96                             &                            0.88                             &                            0.93                             &                            0.96                             &  0.94  \\
 $\Delta({\sigma_{\alpha_{iso}}})$ &                            1.91                             &                            5.05                             &                            -2.20                            &                            -0.84                            &                            0.98                             &  0.98  \\
   $r({\sigma_{\alpha_{iso}}})$    &                            0.94                             &                            0.93                             &                            0.83                             &                            0.87                             &                            0.93                             &  0.90  \\
      $\Delta({\alpha_{ap}})$      &                            -0.05                            &                            -0.05                            &                            0.39                             &                            0.02                             &                            -0.01                            &  0.06  \\
        $r({\alpha_{ap}})$         &                            0.95                             &                            0.96                             &                            0.84                             &                            0.91                             &                            0.96                             &  0.92  \\
 $\Delta({\sigma_{\alpha_{ap}}})$  &                            2.47                             &                            4.27                             &                            -1.22                            &                            -3.04                            &                            -0.11                            &  0.47  \\
    $r({\sigma_{\alpha_{ap}}})$    &                            0.89                             &                            0.89                             &                            0.77                             &                            0.85                             &                            0.92                             &  0.86  \\
         $\Delta({dm+1})$          &                            0.08                             &                            -0.00                            &                            -0.18                            &                            -0.03                            &                            -0.13                            & -0.05  \\
            $r({dm+1})$            &                            0.90                             &                            0.94                             &                            0.77                             &                            0.97                             &                            0.97                             &  0.91  \\
     $\Delta({\sigma_{dm+1}})$     &                            2.49                             &                            4.53                             &                            0.41                             &                            -1.41                            &                            2.00                             &  1.61  \\
       $r({\sigma_{dm+1}})$        &                            0.79                             &                            0.81                             &                            0.82                             &                            0.85                             &                            0.90                             &  0.84  \\
          $\Delta({df})$           &                            0.02                             &                            -0.00                            &                            0.14                             &                            -0.13                            &                            -0.21                            & -0.04  \\
             $r({df})$             &                            0.96                             &                            0.97                             &                            0.93                             &                            0.95                             &                            0.96                             &  0.95  \\
      $\Delta({\sigma_{df}})$      &                            3.77                             &                            5.79                             &                            2.92                             &                            1.27                             &                            4.80                             &  3.71  \\
        $r({\sigma_{df}})$         &                            0.90                             &                            0.90                             &                            0.86                             &                            0.89                             &                            0.93                             &  0.90  \\
        $\Delta({\chi^2})$         &                            -0.11                            &                            1.04                             &                            1.85                             &                            1.57                             &                            0.68                             &  1.01  \\
           $r({\chi^2})$           &                            0.97                             &                            0.97                             &                            0.97                             &                            0.97                             &                            0.97                             &  0.97  \\
\hline
\end{tabular}

    \caption{Comparison statistics for ShapeFit results in configuration space with different samples. Same convention as \cref{tab:bao-avg-all-tracers}.}
    \label{tab:shapefit-avg-all-tracers-rascalc}
\end{table}
\subsubsection{Direct fit}
\begin{figure}
    \centering
    \includegraphics[width=\textwidth]{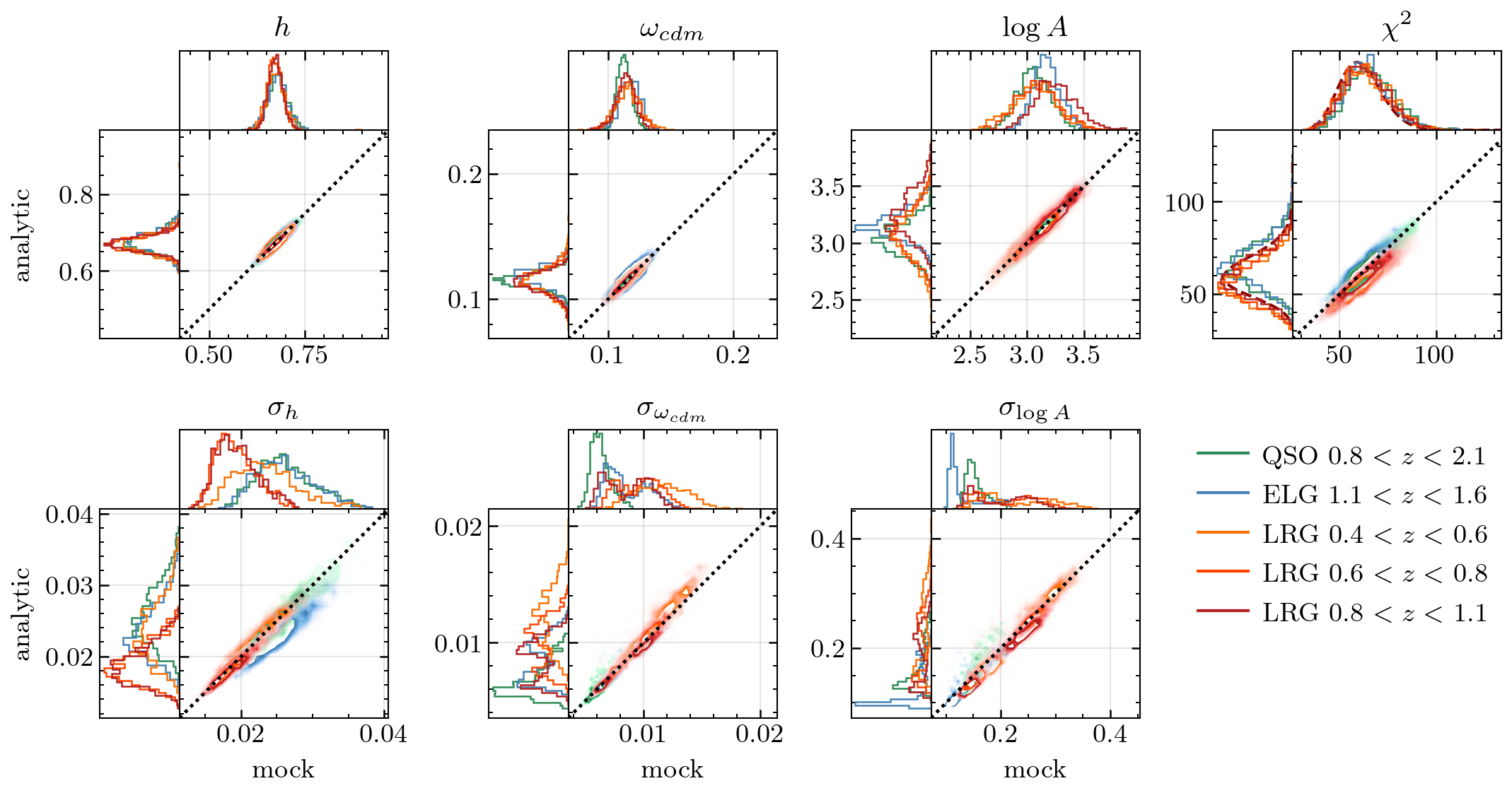}
    \caption{Direct fit comparison for the observed samples. The top row shows the comparison plots for the relevant parameters as well as the $\chi^2$ goodness-of-fit comparison. The bottom row shows the comparison of the error estimates for the parameters.}
    \label{fig:direct-avg-lrg-lowz}
\end{figure}

For completeness, we repeat the same battery of tests using the direct fit approach to Full-Shape measurements. \Cref{fig:direct-avg-lrg-lowz} shows the comparison of cosmological parameters obtained from direct fits with different covariance matrices. While on average there is no significant bias in the parameters themselves, it is worth noticing that the minimiser struggles when fitting some of the realisations in Fourier space, which results in some outliers in the $h$ and $\omega_{cdm}$ estimates. On the other hand, estimates of the errors on the parameters are in general underestimated by the fits with the analytic covariance. The discrepancies range from 7-10\% for $h$ but are otherwise below 5\% level overall. Values individual to each tracer sample can be found in \cref{tab:direct-avg-all-tracers}. We must however highlight that while the rescaled Fourier space analytical covariance can apparently yield error estimates consistent with the sample estimate at this level, the goodness-of-fit is discrepant at a 10\% level for some LRG samples. In fact, these samples are the ones that showed a larger difference in eigenvalues between covariance matrix estimates.

\begin{table}
    \centering
    \rowcolors{1}{white}{mygray}
\begin{tabular}{ccccccc}
\hline
                                   &  \begin{tabular}[c]{@{}c@{}}QSO\\$(0.8, 2.1)$\end{tabular}  &  \begin{tabular}[c]{@{}c@{}}ELG\\$(1.1, 1.6)$\end{tabular}  &  \begin{tabular}[c]{@{}c@{}}LRG\\$(0.4, 0.6)$\end{tabular}  &  \begin{tabular}[c]{@{}c@{}}LRG\\$(0.6, 0.8)$\end{tabular}  &  \begin{tabular}[c]{@{}c@{}}LRG\\$(0.8, 1.1)$\end{tabular}  &  Avg.  \\
\hline
           $\Delta({h})$           &                            0.28                             &                            0.09                             &                            0.09                             &                            0.11                             &                            0.04                             &  0.12  \\
             $r({h})$              &                            0.94                             &                            0.97                             &                            0.88                             &                            0.97                             &                            0.97                             &  0.95  \\
      $\Delta({\sigma_{h}})$       &                            1.63                             &                            10.69                            &                            -1.02                            &                            0.98                             &                            3.33                             &  3.12  \\
         $r({\sigma_{h}})$         &                            0.94                             &                            0.95                             &                            0.95                             &                            0.96                             &                            0.97                             &  0.96  \\
     $\Delta({\omega_{cdm}})$      &                            -2.12                            &                            -0.13                            &                            0.42                             &                            0.38                             &                            0.15                             & -0.26  \\
        $r({\omega_{cdm}})$        &                            0.94                             &                            0.88                             &                            0.94                             &                            0.95                             &                            0.96                             &  0.93  \\
 $\Delta({\sigma_{\omega_{cdm}}})$ &                            -4.75                            &                            1.15                             &                            -5.13                            &                            -2.22                            &                            0.61                             & -2.07  \\
   $r({\sigma_{\omega_{cdm}}})$    &                            0.63                             &                            0.89                             &                            0.89                             &                            0.88                             &                            0.85                             &  0.83  \\
        $\Delta({\log A})$         &                            0.91                             &                            0.23                             &                            -0.11                            &                            -0.00                            &                            0.02                             &  0.21  \\
           $r({\log A})$           &                            0.94                             &                            0.96                             &                            0.94                             &                            0.96                             &                            0.96                             &  0.95  \\
    $\Delta({\sigma_{\log A}})$    &                            -2.16                            &                            3.87                             &                            0.72                             &                            3.41                             &                            6.68                             &  2.51  \\
      $r({\sigma_{\log A}})$       &                            0.82                             &                            0.86                             &                            0.90                             &                            0.89                             &                            0.89                             &  0.87  \\
        $\Delta({\chi^2})$         &                            0.53                             &                            -4.79                            &                            12.26                            &                            10.18                            &                            6.10                             &  4.86  \\
           $r({\chi^2})$           &                            0.96                             &                            0.97                             &                            0.96                             &                            0.96                             &                            0.96                             &  0.96  \\
\hline
\end{tabular}

    \caption{Comparison statistics for direct fit results in Fourier space with different samples. Same convention as \cref{tab:bao-avg-all-tracers,tab:shapefit-avg-all-tracers}.}
    \label{tab:direct-avg-all-tracers}
\end{table}

Analogously to our ShapeFit tests, we perform direct fits in configuration space and find that in this case the results are more consistent between mock and analytic estimates. In particular, we ought to highlight that the largest discrepancy in the error estimates is of 7\% in $\sigma_{\log A}$, which is indeed higher than the average estimate from Fourier space tests. However, goodness-of-fit in configuration space is more consistent between matrix estimates with differences of at most $\sim2\%$. The per-tracer result of the comparison can be found in \cref{tab:direct-avg-all-tracers-rascalc}.

\begin{table}[]
    \centering
    \rowcolors{1}{white}{mygray}
    \begin{tabular}{ccccccc}
\hline
                                   &  \begin{tabular}[c]{@{}c@{}}QSO\\$(0.8, 2.1)$\end{tabular}  &  \begin{tabular}[c]{@{}c@{}}ELG\\$(1.1, 1.6)$\end{tabular}  &  \begin{tabular}[c]{@{}c@{}}LRG\\$(0.4, 0.6)$\end{tabular}  &  \begin{tabular}[c]{@{}c@{}}LRG\\$(0.6, 0.8)$\end{tabular}  &  \begin{tabular}[c]{@{}c@{}}LRG\\$(0.8, 1.1)$\end{tabular}  &  Avg.  \\
\hline
           $\Delta({h})$           &                            0.09                             &                            0.15                             &                            0.01                             &                            0.01                             &                            0.04                             &  0.06  \\
             $r({h})$              &                            0.96                             &                            0.93                             &                            0.86                             &                            0.96                             &                            0.96                             &  0.93  \\
      $\Delta({\sigma_{h}})$       &                            1.30                             &                            5.69                             &                            -1.85                            &                            -1.38                            &                            1.43                             &  1.04  \\
         $r({\sigma_{h}})$         &                            0.95                             &                            0.93                             &                            0.92                             &                            0.96                             &                            0.96                             &  0.94  \\
     $\Delta({\omega_{cdm}})$      &                            -0.04                            &                            0.05                             &                            0.04                             &                            -0.10                            &                            -0.18                            & -0.05  \\
        $r({\omega_{cdm}})$        &                            0.92                             &                            0.89                             &                            0.79                             &                            0.97                             &                            0.97                             &  0.91  \\
 $\Delta({\sigma_{\omega_{cdm}}})$ &                            3.54                             &                            5.41                             &                            -1.55                            &                            -2.15                            &                            2.98                             &  1.65  \\
   $r({\sigma_{\omega_{cdm}}})$    &                            0.84                             &                            0.87                             &                            0.85                             &                            0.89                             &                            0.92                             &  0.87  \\
        $\Delta({\log A})$         &                            -0.10                            &                            -0.07                            &                            0.24                             &                            0.11                             &                            -0.03                            &  0.03  \\
           $r({\log A})$           &                            0.96                             &                            0.89                             &                            0.96                             &                            0.96                             &                            0.96                             &  0.94  \\
    $\Delta({\sigma_{\log A}})$    &                            2.68                             &                            7.44                             &                            3.88                             &                            3.55                             &                            3.70                             &  4.25  \\
      $r({\sigma_{\log A}})$       &                            0.91                             &                            0.87                             &                            0.86                             &                            0.92                             &                            0.93                             &  0.90  \\
        $\Delta({\chi^2})$         &                            0.03                             &                            1.10                             &                            2.18                             &                            1.97                             &                            0.95                             &  1.25  \\
           $r({\chi^2})$           &                            0.97                             &                            0.97                             &                            0.97                             &                            0.97                             &                            0.97                             &  0.97  \\
\hline
\end{tabular}

    \caption{Comparison statistics for direct fit results in configuration space with different samples. Same convention as \cref{tab:bao-avg-all-tracers,tab:shapefit-avg-all-tracers}.}
    \label{tab:direct-avg-all-tracers-rascalc}
\end{table}

\section{Constructing covariance matrix for DESI Year 1 Full Shape analyses}

The tests shown throughout this paper have led us to conclude that the \thecov method of generating analytical covariance matrices for Fourier space clustering does not properly capture the variance in the mocks used, which could be due to a number of factors such as the Fiber Assignment method used in the mocks, the very complex sky footprint of the Year 1 DESI data and how systematics are taken into account (i.e. weighting). 

We refer the reader to Ref. \cite{KP4s8-Alves} for an in-depth study of these phenomena, considering clustering and fibre assignment approximations in our fast simulations.
Ultimately, these findings do not yet allow us to recommend analytic covariances over mock-based ones in Fourier space for the full-shape analyses.

However, mock-based covariance matrices have been found not to represent DESI DR1 perfectly either. The evidence for this comes from the configuration space results and comparisons with RascalC covariance matrices tuned on the data. It was first noticed that configuration-space BAO fits on the data returned $\sim$ 10\% smaller uncertainties and poor goodness-of-fit results when using the mock-based covariance matrix, in comparison to the RascalC covariance matrix tuned on the data \cite{DESI2024.III.KP4}. As demonstrated in this work, results using the mock-based covariance matrix agree well with the RascalC covariance matrix tuned on the mocks. Our conclusion is that the RascalC method is able to infer differences between the data and the mocks that are important for the covariance. Given that the 2-point clustering matches well between the mocks and the data \cite{DESI2024.II.KP3}, potential reasons are discrepancies in 3- and 4-point clustering and the inaccuracies of the Fast Fiber Assign (FFA) algorithm \cite{KP3s6-Bianchi} applied to \ezmock. These results motivated the use of configuration space BAO fits using the \rascalc{} covariance as the primary DESI Y1 BAO measurements.

DESI Y1 full-shape measurements were required to be in Fourier space, due to developments in the modeling and analysis pipeline \cite{DESI2024.V.KP5}. Thus, to account for the apparent difference in covariance between the mocks and the data, we have decided to rescale the mock-based covariance matrices for Full-Shape analysis by a sample-specific factor based on configuration-space comparison with data-tuned \rascalc{} results.
We use the mean ``reduced chi-squared'' measure from Ref.~\cite{Misha2023,KP4s7-Rashkovetskyi}:
\begin{equation} \label{eq:chi2_red-definition}
\chi^2_{\rm red} \qty({\bf C}_R^{-1}, {\bf C}_{\rm s}) \equiv \frac1{N_{\rm bins}} \tr({\bf C}_R^{-1} {\bf C}_{\rm s}),
\end{equation}
where ${\bf C}_R$ is a \rascalc{} covariance matrix tuned on data, ${\bf C}_{\rm s}$ is a configuration-space \ezmock sample covariance (\cref{eq:sample-cov-def}) and $N_{\rm bins}$ is the dimension of both matrices.
If the \rascalc{} covariance matrix describes the distribution of \ezmock clustering measurements, the ``reduced chi-squared'' value is expected to be 1 with a small errorbar \cite{Misha2023,KP4s7-Rashkovetskyi}.
Thus we can bring the mock-based covariance into closer agreement with the \rascalc{} result by multiplying the mock covariance matrix by $1/\chi^2_{\rm red} \qty({\bf C}_R^{-1}, {\bf C}_{\rm s})$.
The ``reduced chi-squared'' values only vary a few per cent depending on the separation range and the number of multipoles \cite{DESI2024.II.KP3}.
The final values of the correction factor vary from 10 to 40\% depending on the sample. The actual values for all tracers and redshift bins can be found in Ref. \cite{DESI2024.V.KP5,DESI2024.II.KP3}. DESI DR1 full shape analysis (\cite{DESI2024.V.KP5}) therefore adopted the sample covariance matrix rescaled with this factor.

\section{Discussion and Conclusion}
\label{sec:discussion-conclusion}

In this work, we have compared analytic estimates of the covariance matrix for two-point clustering measurements in configuration and Fourier space against the corresponding sample covariance. This was achieved first by generating analytic covariance matrices using the mock sample itself, instead of the observed data. In doing this we avoid the assumption that mocks and data are drawn from the same underlying distribution and therefore make our results independent of the mock-generation mechanism. Moreover, we do not limit our tests to visual comparisons between matrix elements, but perform cosmological analyses on the whole sample of mocks to deduce the effects of the differences in covariance matrices in terms of parameter estimates and their uncertainties. In particular, we focus on the fiducial DESI 2024 analysis pipeline, which focuses on 2D Baryon Acoustic Oscillations (BAO) fits using the configuration space two-point function measurements of BAO-reconstructed galaxy samples and Full-Shape fits using Fourier-space measurements of pre-reconstruction samples.

For our BAO-based configuration space tests we have used the \rascalc{} technique and find that parameter constraints obtained with the analytical estimates are consistent with the corresponding parameters obtained when using the sample covariance. In addition, we obtain no significant biases on the error estimates either. These results apply to matrices tuned to individual mock realisations (as would be done for the data) as well as those tuned on the full mock sample. We conclude that given a single clustering measurement (i.e. DESI data), the \rascalc{} covariance generated from it yields good estimates of the errors on cosmological parameters and was used as the covariance matrix of the baseline DESI BAO analysis on DR1 data \citep{DESI2024.III.KP4}.

On the contrary, we find from Fourier-space testing that the mock and analytical covariance matrix estimates show significant differences (around 10\%) not only in an element-to-element comparison but that these large discrepancies induce equally significant differences in the error estimates of cosmological parameters. We have tested both ShapeFit and direct fit Full-Shape analysis techniques and found biases in some of the parameter errors and $\chi^2$ as large as 12\%. We do not expect that these large differences are due only to the covariance given that full-shape analyses are expected to be more challenging than BAO-only fits. However, after testing on full-shape fits in configuration space, we find that the discrepancies are around 7\% for parameter errors and less than 3\% for goodness-of-fit. These tests show that the analytical estimate of the covariance in Fourier space does not yield satisfactory results for Full-Shape analyses when analysing fibre-assigned samples. The inclusion of this effect in configuration space estimates of the covariance is built-in by default through the use of appropriate random catalogues, whereas the inclusion of such effects in Fourier space is not trivial due to the window function associated with these measurements.

Due to the issues observed in Fourier space, coupled with the final sampling technique (i.e. combining BAO and Full-Shape at the data-vector level), the decision was taken to use the mock covariance for the 2024 DESI results. The mock covariance will however be rescaled by a sample-dependent factor that aims at correcting the observed underestimation of variance of the EZmocks when compared to data-based configuration space analytical covariance matrices.

\section*{Data availability}
Data from the plots in this paper is publicly available at \url{https://doi.org/10.5281/zenodo.15120176}

\acknowledgments

DFS acknowledges support from the Swiss National Science Foundation (SNF) "Cosmology with 3D Maps of the Universe" research grant, 200020\_175751 and 200020\_207379. H-JS acknowledges support from the U.S. Department of Energy, Office of Science, Office of High Energy Physics under grant No. DE-SC0023241. 

This material is based upon work supported by the U.S. Department of Energy (DOE), Office of Science, Office of High-Energy Physics, under Contract No. DE–AC02–05CH11231, and by the National Energy Research Scientific Computing Center, a DOE Office of Science User Facility under the same contract. Additional support for DESI was provided by the U.S. National Science Foundation (NSF), Division of Astronomical Sciences under Contract No. AST-0950945 to the NSF’s National Optical-Infrared Astronomy Research Laboratory; the Science and Technology Facilities Council of the United Kingdom; the Gordon and Betty Moore Foundation; the Heising-Simons Foundation; the French Alternative Energies and Atomic Energy Commission (CEA); the National Council of Humanities, Science and Technology of Mexico (CONACYT); the Ministry of Science and Innovation of Spain (MICINN), and by the DESI Member Institutions: \url{https://www.desi.lbl.gov/collaborating-institutions}. Any opinions, findings, and conclusions or recommendations expressed in this material are those of the author(s) and do not necessarily reflect the views of the U. S. National Science Foundation, the U. S. Department of Energy, or any of the listed funding agencies.

The authors are honored to be permitted to conduct scientific research on Iolkam Du’ag (Kitt Peak), a mountain with particular significance to the Tohono O’odham Nation.


\bibliographystyle{JHEP} 
\bibliography{refs, DESI2024} 

\providecommand{\href}[2]{#2}\begingroup\raggedright\begin{thebibliography}{10}

\bibitem{Snowmass2013.Levi}
M.~{Levi}, C.~{Bebek}, T.~{Beers}, R.~{Blum}, R.~{Cahn}, D.~{Eisenstein}
  et~al., \emph{{The DESI Experiment, a whitepaper for Snowmass 2013}},
  {\emph{arXiv e-prints} (2013) arXiv:1308.0847}
  [\href{https://arxiv.org/abs/1308.0847}{{\ttfamily 1308.0847}}].

\bibitem{DESI2016a.Science}
{DESI Collaboration}, A.~{Aghamousa}, J.~{Aguilar}, S.~{Ahlen}, S.~{Alam},
  L.E.~{Allen} et~al., \emph{{The DESI Experiment Part I: Science,Targeting,
  and Survey Design}}, {\emph{arXiv e-prints} (2016) arXiv:1611.00036}
  [\href{https://arxiv.org/abs/1611.00036}{{\ttfamily 1611.00036}}].

\bibitem{DESI2016b.Instr}
{DESI Collaboration}, A.~{Aghamousa}, J.~{Aguilar}, S.~{Ahlen}, S.~{Alam},
  L.E.~{Allen} et~al., \emph{{The DESI Experiment Part II: Instrument Design}},
  {\emph{arXiv e-prints} (2016) arXiv:1611.00037}
  [\href{https://arxiv.org/abs/1611.00037}{{\ttfamily 1611.00037}}].

\bibitem{DESI2022.KP1.Instr}
{DESI Collaboration}, B.~{Abareshi}, J.~{Aguilar}, S.~{Ahlen}, S.~{Alam},
  D.M.~{Alexander} et~al., \emph{{Overview of the Instrumentation for the Dark
  Energy Spectroscopic Instrument}},
  \href{https://doi.org/10.3847/1538-3881/ac882b}{\emph{\aj} {\bfseries 164}
  (2022) 207} [\href{https://arxiv.org/abs/2205.10939}{{\ttfamily
  2205.10939}}].

\bibitem{FocalPlane.Silber.2023}
J.H.~{Silber}, P.~{Fagrelius}, K.~{Fanning}, M.~{Schubnell}, J.N.~{Aguilar},
  S.~{Ahlen} et~al., \emph{{The Robotic Multiobject Focal Plane System of the
  Dark Energy Spectroscopic Instrument (DESI)}},
  \href{https://doi.org/10.3847/1538-3881/ac9ab1}{\emph{\aj} {\bfseries 165}
  (2023) 9} [\href{https://arxiv.org/abs/2205.09014}{{\ttfamily 2205.09014}}].

\bibitem{Corrector.Miller.2023}
T.N.~{Miller}, P.~{Doel}, G.~{Gutierrez}, R.~{Besuner}, D.~{Brooks}, G.~{Gallo}
  et~al., \emph{{The Optical Corrector for the Dark Energy Spectroscopic
  Instrument}}, \href{https://doi.org/10.3847/1538-3881/ad45fe}{\emph{\aj}
  {\bfseries 168} (2024) 95}
  [\href{https://arxiv.org/abs/2306.06310}{{\ttfamily 2306.06310}}].

\bibitem{Spectro.Pipeline.Guy.2023}
J.~{Guy}, S.~{Bailey}, A.~{Kremin}, S.~{Alam}, D.M.~{Alexander}, C.~{Allende
  Prieto} et~al., \emph{{The Spectroscopic Data Processing Pipeline for the
  Dark Energy Spectroscopic Instrument}},
  \href{https://doi.org/10.3847/1538-3881/acb212}{\emph{\aj} {\bfseries 165}
  (2023) 144} [\href{https://arxiv.org/abs/2209.14482}{{\ttfamily
  2209.14482}}].

\bibitem{SurveyOps.Schlafly.2023}
E.F.~{Schlafly}, D.~{Kirkby}, D.J.~{Schlegel}, A.D.~{Myers}, A.~{Raichoor},
  K.~{Dawson} et~al., \emph{{Survey Operations for the Dark Energy
  Spectroscopic Instrument}},
  \href{https://doi.org/10.3847/1538-3881/ad0832}{\emph{\aj} {\bfseries 166}
  (2023) 259} [\href{https://arxiv.org/abs/2306.06309}{{\ttfamily
  2306.06309}}].

\bibitem{Levi2019}
M.~{Levi}, L.E.~{Allen}, A.~{Raichoor}, C.~{Baltay}, S.~{BenZvi}, F.~{Beutler}
  et~al., \emph{{The Dark Energy Spectroscopic Instrument (DESI)}},  in
  \emph{Bulletin of the American Astronomical Society}, vol.~51, p.~57, Sept.,
  2019, \href{https://doi.org/10.48550/arXiv.1907.10688}{DOI}
  [\href{https://arxiv.org/abs/1907.10688}{{\ttfamily 1907.10688}}].

\bibitem{DESI2023a.KP1.SV}
{DESI Collaboration}, A.G.~{Adame}, J.~{Aguilar}, S.~{Ahlen}, S.~{Alam},
  G.~{Aldering} et~al., \emph{{Validation of the Scientific Program for the
  Dark Energy Spectroscopic Instrument}},
  \href{https://doi.org/10.3847/1538-3881/ad0b08}{\emph{\aj} {\bfseries 167}
  (2024) 62} [\href{https://arxiv.org/abs/2306.06307}{{\ttfamily 2306.06307}}].

\bibitem{DESI2023b.KP1.EDR}
{DESI Collaboration}, A.G.~{Adame}, J.~{Aguilar}, S.~{Ahlen}, S.~{Alam},
  G.~{Aldering} et~al., \emph{{The Early Data Release of the Dark Energy
  Spectroscopic Instrument}},
  \href{https://doi.org/10.3847/1538-3881/ad3217}{\emph{\aj} {\bfseries 168}
  (2024) 58} [\href{https://arxiv.org/abs/2306.06308}{{\ttfamily 2306.06308}}].

\bibitem{Eisenstein2011}
D.J.~{Eisenstein}, D.H.~{Weinberg}, E.~{Agol}, H.~{Aihara}, C.~{Allende
  Prieto}, S.F.~{Anderson} et~al., \emph{{SDSS-III: Massive Spectroscopic
  Surveys of the Distant Universe, the Milky Way, and Extra-Solar Planetary
  Systems}}, \href{https://doi.org/10.1088/0004-6256/142/3/72}{\emph{\aj}
  {\bfseries 142} (2011) 72} [\href{https://arxiv.org/abs/1101.1529}{{\ttfamily
  1101.1529}}].

\bibitem{Dawson2013}
K.S.~{Dawson}, D.J.~{Schlegel}, C.P.~{Ahn}, S.F.~{Anderson}, {\'E}.~{Aubourg},
  S.~{Bailey} et~al., \emph{{The Baryon Oscillation Spectroscopic Survey of
  SDSS-III}}, \href{https://doi.org/10.1088/0004-6256/145/1/10}{\emph{\aj}
  {\bfseries 145} (2013) 10} [\href{https://arxiv.org/abs/1208.0022}{{\ttfamily
  1208.0022}}].

\bibitem{Cole05}
S.~{Cole}, W.J.~{Percival}, J.A.~{Peacock}, P.~{Norberg}, C.M.~{Baugh},
  C.S.~{Frenk} et~al., \emph{{The 2dF Galaxy Redshift Survey: power-spectrum
  analysis of the final data set and cosmological implications}},
  \href{https://doi.org/10.1111/j.1365-2966.2005.09318.x}{\emph{\mnras}
  {\bfseries 362} (2005) 505}
  [\href{https://arxiv.org/abs/astro-ph/0501174}{{\ttfamily
  astro-ph/0501174}}].

\bibitem{Eisenstein05}
D.J.~{Eisenstein}, I.~{Zehavi}, D.W.~{Hogg}, R.~{Scoccimarro}, M.R.~{Blanton},
  R.C.~{Nichol} et~al., \emph{{Detection of the Baryon Acoustic Peak in the
  Large-Scale Correlation Function of SDSS Luminous Red Galaxies}},
  \href{https://doi.org/10.1086/466512}{\emph{\apj} {\bfseries 633} (2005) 560}
  [\href{https://arxiv.org/abs/astro-ph/0501171}{{\ttfamily
  astro-ph/0501171}}].

\bibitem{Beutler11}
F.~{Beutler}, C.~{Blake}, M.~{Colless}, D.H.~{Jones}, L.~{Staveley-Smith},
  L.~{Campbell} et~al., \emph{{The 6dF Galaxy Survey: baryon acoustic
  oscillations and the local Hubble constant}},
  \href{https://doi.org/10.1111/j.1365-2966.2011.19250.x}{\emph{\mnras}
  {\bfseries 416} (2011) 3017}
  [\href{https://arxiv.org/abs/1106.3366}{{\ttfamily 1106.3366}}].

\bibitem{Alam2021}
S.~{Alam}, M.~{Aubert}, S.~{Avila}, C.~{Balland}, J.E.~{Bautista},
  M.A.~{Bershady} et~al., \emph{{Completed SDSS-IV extended Baryon Oscillation
  Spectroscopic Survey: Cosmological implications from two decades of
  spectroscopic surveys at the Apache Point Observatory}},
  \href{https://doi.org/10.1103/PhysRevD.103.083533}{\emph{\prd} {\bfseries
  103} (2021) 083533} [\href{https://arxiv.org/abs/2007.08991}{{\ttfamily
  2007.08991}}].

\bibitem{Brieden21}
S.~{Brieden}, H.~{Gil-Mar{\'\i}n} and L.~{Verde}, \emph{{ShapeFit: extracting
  the power spectrum shape information in galaxy surveys beyond BAO and RSD}},
  \href{https://doi.org/10.1088/1475-7516/2021/12/054}{\emph{\jcap} {\bfseries
  2021} (2021) 054} [\href{https://arxiv.org/abs/2106.07641}{{\ttfamily
  2106.07641}}].

\bibitem{DESI2024.V.KP5}
{DESI Collaboration}, A.G.~{Adame}, J.~{Aguilar}, S.~{Ahlen}, S.~{Alam},
  D.M.~{Alexander} et~al., \emph{{DESI 2024 V: Full-Shape Galaxy Clustering
  from Galaxies and Quasars}},
  \href{https://doi.org/10.48550/arXiv.2411.12021}{\emph{arXiv e-prints} (2024)
  arXiv:2411.12021} [\href{https://arxiv.org/abs/2411.12021}{{\ttfamily
  2411.12021}}].

\bibitem{Maus23}
M.~{Maus}, S.-F.~{Chen} and M.~{White}, \emph{{A comparison of template vs.
  direct model fitting for redshift-space distortions in BOSS}},
  \href{https://doi.org/10.1088/1475-7516/2023/06/005}{\emph{\jcap} {\bfseries
  2023} (2023) 005} [\href{https://arxiv.org/abs/2302.07430}{{\ttfamily
  2302.07430}}].

\bibitem{DESI2024.III.KP4}
{DESI Collaboration}, A.G.~Adame, J.~Aguilar, S.~Ahlen, S.~Alam, D.M.~Alexander
  et~al., \emph{{DESI 2024 III: Baryon Acoustic Oscillations from Galaxies and
  Quasars}}, \href{https://doi.org/10.48550/arXiv.2404.03000}{\emph{arXiv
  e-prints} (2024) arXiv:2404.03000}
  [\href{https://arxiv.org/abs/2404.03000}{{\ttfamily 2404.03000}}].

\bibitem{DESI2024.I.DR1}
{DESI Collaboration}, M.~{Abdul-Karim}, A.G.~{Adame}, D.~{Aguado},
  J.~{Aguilar}, S.~{Ahlen} et~al., \emph{{Data Release 1 of the Dark Energy
  Spectroscopic Instrument}},
  \href{https://doi.org/10.48550/arXiv.2503.14745}{\emph{arXiv e-prints} (2025)
  arXiv:2503.14745} [\href{https://arxiv.org/abs/2503.14745}{{\ttfamily
  2503.14745}}].

\bibitem{DESI2024.II.KP3}
{DESI Collaboration}, A.G.~{Adame}, J.~{Aguilar}, S.~{Ahlen}, S.~{Alam},
  D.M.~{Alexander} et~al., \emph{{DESI 2024 II: Sample Definitions,
  Characteristics, and Two-point Clustering Statistics}},
  \href{https://doi.org/10.48550/arXiv.2411.12020}{\emph{arXiv e-prints} (2024)
  arXiv:2411.12020} [\href{https://arxiv.org/abs/2411.12020}{{\ttfamily
  2411.12020}}].

\bibitem{DESI2024.IV.KP6}
{DESI Collaboration}, A.G.~{Adame}, J.~{Aguilar}, S.~{Ahlen}, S.~{Alam},
  D.M.~{Alexander} et~al., \emph{{DESI 2024 IV: Baryon Acoustic Oscillations
  from the Lyman alpha forest}},
  \href{https://doi.org/10.1088/1475-7516/2025/01/124}{\emph{\jcap} {\bfseries
  2025} (2025) 124} [\href{https://arxiv.org/abs/2404.03001}{{\ttfamily
  2404.03001}}].

\bibitem{DESI2024.VI.KP7A}
{DESI Collaboration}, A.G.~{Adame}, J.~{Aguilar}, S.~{Ahlen}, S.~{Alam},
  D.M.~{Alexander} et~al., \emph{{DESI 2024 VI: cosmological constraints from
  the measurements of baryon acoustic oscillations}},
  \href{https://doi.org/10.1088/1475-7516/2025/02/021}{\emph{\jcap} {\bfseries
  2025} (2025) 021} [\href{https://arxiv.org/abs/2404.03002}{{\ttfamily
  2404.03002}}].

\bibitem{DESI2024.VII.KP7B}
{DESI Collaboration}, A.G.~{Adame}, J.~{Aguilar}, S.~{Ahlen}, S.~{Alam},
  D.M.~{Alexander} et~al., \emph{{DESI 2024 VII: Cosmological Constraints from
  the Full-Shape Modeling of Clustering Measurements}},
  \href{https://doi.org/10.48550/arXiv.2411.12022}{\emph{arXiv e-prints} (2024)
  arXiv:2411.12022} [\href{https://arxiv.org/abs/2411.12022}{{\ttfamily
  2411.12022}}].

\bibitem{Chuang2015}
C.-H.~{Chuang}, C.~{Zhao}, F.~{Prada}, E.~{Munari}, S.~{Avila}, A.~{Izard}
  et~al., \emph{{nIFTy cosmology: Galaxy/halo mock catalogue comparison project
  on clustering statistics}},
  \href{https://doi.org/10.1093/mnras/stv1289}{\emph{\mnras} {\bfseries 452}
  (2015) 686} [\href{https://arxiv.org/abs/1412.7729}{{\ttfamily 1412.7729}}].

\bibitem{Lippich2019}
M.~{Lippich}, A.G.~{S{\'a}nchez}, M.~{Colavincenzo}, E.~{Sefusatti},
  P.~{Monaco}, L.~{Blot} et~al., \emph{{Comparing approximate methods for mock
  catalogues and covariance matrices - I. Correlation function}},
  \href{https://doi.org/10.1093/mnras/sty2757}{\emph{\mnras} {\bfseries 482}
  (2019) 1786} [\href{https://arxiv.org/abs/1806.09477}{{\ttfamily
  1806.09477}}].

\bibitem{Blot2019}
L.~{Blot}, M.~{Crocce}, E.~{Sefusatti}, M.~{Lippich}, A.G.~{S{\'a}nchez},
  M.~{Colavincenzo} et~al., \emph{{Comparing approximate methods for mock
  catalogues and covariance matrices II: power spectrum multipoles}},
  \href{https://doi.org/10.1093/mnras/stz507}{\emph{\mnras} {\bfseries 485}
  (2019) 2806} [\href{https://arxiv.org/abs/1806.09497}{{\ttfamily
  1806.09497}}].

\bibitem{Colavincenzo2019}
M.~{Colavincenzo}, E.~{Sefusatti}, P.~{Monaco}, L.~{Blot}, M.~{Crocce},
  M.~{Lippich} et~al., \emph{{Comparing approximate methods for mock catalogues
  and covariance matrices - III: bispectrum}},
  \href{https://doi.org/10.1093/mnras/sty2964}{\emph{\mnras} {\bfseries 482}
  (2019) 4883} [\href{https://arxiv.org/abs/1806.09499}{{\ttfamily
  1806.09499}}].

\bibitem{Chuang2015EZmock}
C.-H.~{Chuang}, F.-S.~{Kitaura}, F.~{Prada}, C.~{Zhao} and G.~{Yepes},
  \emph{{EZmocks: extending the Zel'dovich approximation to generate mock
  galaxy catalogues with accurate clustering statistics}},
  \href{https://doi.org/10.1093/mnras/stu2301}{\emph{\mnras} {\bfseries 446}
  (2015) 2621} [\href{https://arxiv.org/abs/1409.1124}{{\ttfamily 1409.1124}}].

\bibitem{Zhao2021}
C.~{Zhao}, C.-H.~{Chuang}, J.~{Bautista}, A.~{de Mattia}, A.~{Raichoor},
  A.J.~{Ross} et~al., \emph{{The completed SDSS-IV extended Baryon Oscillation
  Spectroscopic Survey: 1000 multi-tracer mock catalogues with redshift
  evolution and systematics for galaxies and quasars of the final data
  release}}, \href{https://doi.org/10.1093/mnras/stab510}{\emph{\mnras}
  {\bfseries 503} (2021) 1149}
  [\href{https://arxiv.org/abs/2007.08997}{{\ttfamily 2007.08997}}].

\bibitem{Zeldovich1970}
Y.B.~{Zel'dovich}, \emph{{Gravitational instability: An approximate theory for
  large density perturbations.}}, {\emph{\aap} {\bfseries 5} (1970) 84}.

\bibitem{KP3s8-Zhao}
{C.~Zhao et al.}, \emph{{Mock catalogues with survey realism for the DESI
  DR1}}, {\emph{in preparation} (2025) }.

\bibitem{Yuan2024}
S.~{Yuan}, H.~{Zhang}, A.J.~{Ross}, J.~{Donald-McCann}, B.~{Hadzhiyska},
  R.H.~{Wechsler} et~al., \emph{{The DESI one-per cent survey: exploring the
  halo occupation distribution of luminous red galaxies and quasi-stellar
  objects with ABACUSSUMMIT}},
  \href{https://doi.org/10.1093/mnras/stae359}{\emph{\mnras} {\bfseries 530}
  (2024) 947} [\href{https://arxiv.org/abs/2306.06314}{{\ttfamily
  2306.06314}}].

\bibitem{antoine2023}
A.~{Rocher}, V.~{Ruhlmann-Kleider}, E.~{Burtin}, S.~{Yuan}, A.~{de Mattia},
  A.J.~{Ross} et~al., \emph{{The DESI One-Percent survey: exploring the Halo
  Occupation Distribution of Emission Line Galaxies with ABACUSSUMMIT
  simulations}},
  \href{https://doi.org/10.1088/1475-7516/2023/10/016}{\emph{\jcap} {\bfseries
  2023} (2023) 016} [\href{https://arxiv.org/abs/2306.06319}{{\ttfamily
  2306.06319}}].

\bibitem{KP3s11-Sikandar}
{M.~M.~S~Hanif et al.}, \emph{{Fast Fiber Assign: Emulating fiber assignment
  effects for realistic DESI catalogs}}, {\emph{in preparation} (2025) }.

\bibitem{Hartlap2007}
J.~{Hartlap}, P.~{Simon} and P.~{Schneider}, \emph{{Why your model parameter
  confidences might be too optimistic. Unbiased estimation of the inverse
  covariance matrix}},
  \href{https://doi.org/10.1051/0004-6361:20066170}{\emph{\aap} {\bfseries 464}
  (2007) 399} [\href{https://arxiv.org/abs/astro-ph/0608064}{{\ttfamily
  astro-ph/0608064}}].

\bibitem{Percival2014}
W.J.~{Percival}, A.J.~{Ross}, A.G.~{S{\'a}nchez}, L.~{Samushia}, A.~{Burden},
  R.~{Crittenden} et~al., \emph{{The clustering of Galaxies in the SDSS-III
  Baryon Oscillation Spectroscopic Survey: including covariance matrix
  errors}}, \href{https://doi.org/10.1093/mnras/stu112}{\emph{\mnras}
  {\bfseries 439} (2014) 2531}
  [\href{https://arxiv.org/abs/1312.4841}{{\ttfamily 1312.4841}}].

\bibitem{Percival2022}
W.J.~{Percival}, O.~{Friedrich}, E.~{Sellentin} and A.~{Heavens},
  \emph{{Matching Bayesian and frequentist coverage probabilities when using an
  approximate data covariance matrix}},
  \href{https://doi.org/10.1093/mnras/stab3540}{\emph{\mnras} {\bfseries 510}
  (2022) 3207} [\href{https://arxiv.org/abs/2108.10402}{{\ttfamily
  2108.10402}}].

\bibitem{rascal}
R.~{O'Connell}, D.~{Eisenstein}, M.~{Vargas}, S.~{Ho} and N.~{Padmanabhan},
  \emph{{Large covariance matrices: smooth models from the two-point
  correlation function}},
  \href{https://doi.org/10.1093/mnras/stw1821}{\emph{\mnras} {\bfseries 462}
  (2016) 2681} [\href{https://arxiv.org/abs/1510.01740}{{\ttfamily
  1510.01740}}].

\bibitem{rascal-jackknife}
R.~{O'Connell} and D.J.~{Eisenstein}, \emph{{Large covariance matrices:
  accurate models without mocks}},
  \href{https://doi.org/10.1093/mnras/stz1359}{\emph{\mnras} {\bfseries 487}
  (2019) 2701} [\href{https://arxiv.org/abs/1808.05978}{{\ttfamily
  1808.05978}}].

\bibitem{rascalC}
O.H.E.~{Philcox}, D.J.~{Eisenstein}, R.~{O'Connell} and A.~{Wiegand},
  \emph{{RASCALC: a jackknife approach to estimating single- and multitracer
  galaxy covariance matrices}},
  \href{https://doi.org/10.1093/mnras/stz3218}{\emph{\mnras} {\bfseries 491}
  (2020) 3290} [\href{https://arxiv.org/abs/1904.11070}{{\ttfamily
  1904.11070}}].

\bibitem{rascalC-legendre-3}
O.H.E.~{Philcox} and D.J.~{Eisenstein}, \emph{{Estimating covariance matrices
  for two- and three-point correlation function moments in Arbitrary Survey
  Geometries}}, \href{https://doi.org/10.1093/mnras/stz2896}{\emph{\mnras}
  {\bfseries 490} (2019) 5931}
  [\href{https://arxiv.org/abs/1910.04764}{{\ttfamily 1910.04764}}].

\bibitem{Moon2023}
J.~{Moon}, D.~{Valcin}, M.~{Rashkovetskyi}, C.~{Saulder}, K.~{Dawson}, A.~{de
  Mattia} et~al., ``First detection of the bao signal from early desi data.''
  2023.

\bibitem{Misha2023}
M.~{Rashkovetskyi}, D.J.~{Eisenstein}, J.N.~{Aguilar}, D.~{Brooks},
  T.~{Claybaugh}, S.~{Cole} et~al., \emph{{Validation of semi-analytical,
  semi-empirical covariance matrices for two-point correlation function for
  early DESI data}},
  \href{https://doi.org/10.1093/mnras/stad2078}{\emph{\mnras} {\bfseries 524}
  (2023) 3894} [\href{https://arxiv.org/abs/2306.06320}{{\ttfamily
  2306.06320}}].

\bibitem{KP4s7-Rashkovetskyi}
M.~{Rashkovetskyi}, D.~{Forero-S{\'a}nchez}, A.~{de Mattia}, D.J.~{Eisenstein},
  N.~{Padmanabhan}, H.~{Seo} et~al., \emph{{Semi-analytical covariance matrices
  for two-point correlation function for DESI 2024 data}},
  \href{https://doi.org/10.1088/1475-7516/2025/01/145}{\emph{\jcap} {\bfseries
  2025} (2025) 145} [\href{https://arxiv.org/abs/2404.03007}{{\ttfamily
  2404.03007}}].

\bibitem{Wadekar2019}
D.~Wadekar and R.~Scoccimarro, \emph{{Galaxy power spectrum multipoles
  covariance in perturbation theory}},
  \href{https://doi.org/10.1103/PhysRevD.102.123517}{\emph{Phys. Rev. D}
  {\bfseries 102} (2020) 123517}
  [\href{https://arxiv.org/abs/1910.02914}{{\ttfamily 1910.02914}}].

\bibitem{KP4s8-Alves}
{O.~Alves et al.}, \emph{{Analytical covariance matrices of DESI galaxy power
  spectra}}, {\emph{in preparation} (2025) }.

\bibitem{Minuit}
M.~Hatlo, F.~James, P.~Mato, L.~Moneta, M.~Winkler and A.~Zsenei,
  \emph{Developments of mathematical software libraries for the lhc
  experiments},  in \emph{IEEE Symposium Conference Record Nuclear Science
  2004.}, vol.~4, pp.~2091--2094 Vol. 4, 2004,
  \href{https://doi.org/10.1109/NSSMIC.2004.1462675}{DOI}.

\bibitem{KP4s2-Chen}
S.F.~{Chen}, C.~{Howlett}, M.~{White}, P.~{McDonald}, A.J.~{Ross}, H.J.~{Seo}
  et~al., \emph{{Baryon acoustic oscillation theory and modelling systematics
  for the DESI 2024 results}},
  \href{https://doi.org/10.1093/mnras/stae2090}{\emph{\mnras} {\bfseries 534}
  (2024) 544} [\href{https://arxiv.org/abs/2402.14070}{{\ttfamily
  2402.14070}}].

\bibitem{Burden2015}
A.~{Burden}, W.J.~{Percival} and C.~{Howlett}, \emph{{Reconstruction in Fourier
  space}}, \href{https://doi.org/10.1093/mnras/stv1581}{\emph{\mnras}
  {\bfseries 453} (2015) 456}
  [\href{https://arxiv.org/abs/1504.02591}{{\ttfamily 1504.02591}}].

\bibitem{KP4s3-Chen}
X.~{Chen}, Z.~{Ding}, E.~{Paillas}, S.~{Nadathur}, H.~{Seo}, S.~{Chen} et~al.,
  \emph{{Extensive analysis of reconstruction algorithms for DESI 2024 baryon
  acoustic oscillations}},
  \href{https://doi.org/10.48550/arXiv.2411.19738}{\emph{arXiv e-prints} (2024)
  arXiv:2411.19738} [\href{https://arxiv.org/abs/2411.19738}{{\ttfamily
  2411.19738}}].

\bibitem{KP4s4-Paillas}
E.~{Paillas}, Z.~{Ding}, X.~{Chen}, H.~{Seo}, N.~{Padmanabhan}, A.~{de Mattia}
  et~al., \emph{{Optimal reconstruction of baryon acoustic oscillations for
  DESI 2024}},
  \href{https://doi.org/10.1088/1475-7516/2025/01/142}{\emph{\jcap} {\bfseries
  2025} (2025) 142} [\href{https://arxiv.org/abs/2404.03005}{{\ttfamily
  2404.03005}}].

\bibitem{White2015}
M.~{White}, \emph{{Reconstruction within the Zeldovich approximation}},
  \href{https://doi.org/10.1093/mnras/stv842}{\emph{\mnras} {\bfseries 450}
  (2015) 3822} [\href{https://arxiv.org/abs/1504.03677}{{\ttfamily
  1504.03677}}].

\bibitem{Xu2012}
X.~{Xu}, N.~{Padmanabhan}, D.J.~{Eisenstein}, K.T.~{Mehta} and A.J.~{Cuesta},
  \emph{{A 2 per cent distance to z = 0.35 by reconstructing baryon acoustic
  oscillations - II. Fitting techniques}},
  \href{https://doi.org/10.1111/j.1365-2966.2012.21573.x}{\emph{\mnras}
  {\bfseries 427} (2012) 2146}
  [\href{https://arxiv.org/abs/1202.0091}{{\ttfamily 1202.0091}}].

\bibitem{Chen2020}
S.-F.~{Chen}, Z.~{Vlah} and M.~{White}, \emph{{Consistent modeling of velocity
  statistics and redshift-space distortions in one-loop perturbation theory}},
  \href{https://doi.org/10.1088/1475-7516/2020/07/062}{\emph{\jcap} {\bfseries
  2020} (2020) 062} [\href{https://arxiv.org/abs/2005.00523}{{\ttfamily
  2005.00523}}].

\bibitem{Chen2021}
S.-F.~{Chen}, Z.~{Vlah}, E.~{Castorina} and M.~{White}, \emph{{Redshift-space
  distortions in Lagrangian perturbation theory}},
  \href{https://doi.org/10.1088/1475-7516/2021/03/100}{\emph{\jcap} {\bfseries
  2021} (2021) 100} [\href{https://arxiv.org/abs/2012.04636}{{\ttfamily
  2012.04636}}].

\bibitem{KP5s2-Maus}
M.~{Maus}, S.~{Chen}, M.~{White}, J.~{Aguilar}, S.~{Ahlen}, A.~{Aviles} et~al.,
  \emph{{An analysis of parameter compression and Full-Modeling techniques with
  Velocileptors for DESI 2024 and beyond}},
  \href{https://doi.org/10.1088/1475-7516/2025/01/138}{\emph{\jcap} {\bfseries
  2025} (2025) 138} [\href{https://arxiv.org/abs/2404.07312}{{\ttfamily
  2404.07312}}].

\bibitem{Noriega2022}
H.E.~Noriega, A.~Aviles, S.~Fromenteau and M.~Vargas-Maga\~na, \emph{{Fast
  computation of non-linear power spectrum in cosmologies with massive
  neutrinos}}, \href{https://doi.org/10.1088/1475-7516/2022/11/038}{\emph{JCAP}
  {\bfseries 11} (2022) 038}
  [\href{https://arxiv.org/abs/2208.02791}{{\ttfamily 2208.02791}}].

\bibitem{KP5s3-Noriega}
H.E.~{Noriega}, A.~{Aviles}, H.~{Gil-Mar{\'\i}n}, S.~{Ramirez-Solano},
  S.~{Fromenteau}, M.~{Vargas-Maga{\~n}a} et~al., \emph{{Comparing Compressed
  and Full-Modeling analyses with FOLPS: implications for DESI 2024 and
  beyond}}, \href{https://doi.org/10.1088/1475-7516/2025/01/136}{\emph{\jcap}
  {\bfseries 2025} (2025) 136}
  [\href{https://arxiv.org/abs/2404.07269}{{\ttfamily 2404.07269}}].

\bibitem{CLASS}
D.~{Blas}, J.~{Lesgourgues} and T.~{Tram}, \emph{{The Cosmic Linear Anisotropy
  Solving System (CLASS). Part II: Approximation schemes}},
  \href{https://doi.org/10.1088/1475-7516/2011/07/034}{\emph{\jcap} {\bfseries
  7} (2011) 034} [\href{https://arxiv.org/abs/1104.2933}{{\ttfamily
  1104.2933}}].

\bibitem{CAMB}
A.~Lewis, A.~Challinor and A.~Lasenby, \emph{{Efficient computation of CMB
  anisotropies in closed FRW models}},
  \href{https://doi.org/10.1086/309179}{\emph{\apj} {\bfseries 538} (2000) 473}
  [\href{https://arxiv.org/abs/astro-ph/9911177}{{\ttfamily
  astro-ph/9911177}}].

\bibitem{Howlett2012}
C.~{Howlett}, A.~{Lewis}, A.~{Hall} and A.~{Challinor}, \emph{{CMB power
  spectrum parameter degeneracies in the era of precision cosmology}},
  \href{https://doi.org/10.1088/1475-7516/2012/04/027}{\emph{\jcap} {\bfseries
  2012} (2012) 027} [\href{https://arxiv.org/abs/1201.3654}{{\ttfamily
  1201.3654}}].

\bibitem{KP5s1-Maus}
M.~{Maus}, Y.~{Lai}, H.E.~{Noriega}, S.~{Ramirez-Solano}, A.~{Aviles},
  S.~{Chen} et~al., \emph{{A comparison of effective field theory models of
  redshift space galaxy power spectra for DESI 2024 and future surveys}},
  \href{https://doi.org/10.1088/1475-7516/2025/01/134}{\emph{\jcap} {\bfseries
  2025} (2025) 134} [\href{https://arxiv.org/abs/2404.07272}{{\ttfamily
  2404.07272}}].

\bibitem{Beutler2021}
F.~{Beutler} and P.~{McDonald}, \emph{{Unified galaxy power spectrum
  measurements from 6dFGS, BOSS, and eBOSS}},
  \href{https://doi.org/10.1088/1475-7516/2021/11/031}{\emph{\jcap} {\bfseries
  2021} (2021) 031} [\href{https://arxiv.org/abs/2106.06324}{{\ttfamily
  2106.06324}}].

\bibitem{KP3s6-Bianchi}
D.~{Bianchi}, M.M.S.~{Hanif}, A.~{Carnero Rosell}, J.~{Lasker}, A.J.~{Ross},
  M.~{Pinon} et~al., \emph{{Characterization of DESI fiber assignment
  incompleteness effect on 2-point clustering and mitigation methods for DR1
  analysis}}, \href{https://doi.org/10.48550/arXiv.2411.12025}{\emph{arXiv
  e-prints} (2024) arXiv:2411.12025}
  [\href{https://arxiv.org/abs/2411.12025}{{\ttfamily 2411.12025}}].

\end{thebibliography}\endgroup
\end{document}